\documentclass[prl, twocolumn,aps,preprintnumbers,letterpaper,floatfix,showpacs]{revtex4}
\usepackage{bm, graphicx, color}
\usepackage{amsmath,amssymb,amsfonts}
\usepackage{wasysym, hyperref,subfigure}

\newcommand{\ddd}{\displaystyle}
\newcommand{\dd}{\partial}
\newcommand{\half}{\frac{1}{2}}
\newcommand{\LL}[2]{\Lambda_{#1, #2}}
\newcommand{\KK}{\kappa}

\newcommand{\hide}[1]{ }
\newcommand{\vv}{v_g}
\newcommand{\TT}{{\bar T_e}}

\begin{document}

\title{Extracting dark matter signatures from atomic clock stability measurements}

\author{Tigran Kalaydzhyan}\email{tigran.kalaydzhyan@jpl.nasa.gov}
\affiliation{Jet Propulsion Laboratory, California Institute of Technology,\\ 4800 Oak Grove Dr, MS 298, Pasadena, CA 91109, U.S.A.}
\author{Nan Yu}
\affiliation{Jet Propulsion Laboratory, California Institute of Technology,\\ 4800 Oak Grove Dr, MS 298, Pasadena, CA 91109, U.S.A.}

\date{\today}

\begin{abstract}
We analyze possible effects of the dark matter environment on the atomic clock stability measurements. The dark matter is assumed to exist in a form of waves of ultralight scalar fields or in a form of topological defects (monopoles and strings). We identify dark matter signal signatures in clock Allan deviation plots that can be used to constrain the dark matter coupling to the Standard Model fields. The existing data on the Al${}^+$/Hg${}^+$ clock comparison are used to put new limits on the dilaton dark matter in the region of masses $m_\phi > 10^{-15}\mathrm{eV}$. We also estimate the sensitivities of future atomic clock experiments in space, including the cesium microwave and strontium optical clocks aboard the International Space Station, as well as a potential nuclear clock. These experiments are expected to put new limits on the topological dark matter in the range of masses  $10^{-10}\mathrm{eV} < m_\phi < 10^{-6}\mathrm{eV}$.

\end{abstract}


\maketitle

\section{Introduction}
Despite an overwhelming amount of cosmological and astrophysical data suggesting the existence of dark matter (DM), there is no confirmed direct detection of DM particles or fields up to the date, see Ref.~\cite{Olive:2016xmw} for review. 
One of the main obstacles for the direct detection is an enormous space of allowed parameters describing fundametal properties of DM. The mass of the DM field is allowed to span a vast range of values: from the inverse sizes of dwarf galaxies ($10^{-24}\, \mathrm{eV}$, see, e.g., Ref.~\cite{Marsh:2015xka}) all the way to the Planck mass ($10^{28}\, \mathrm{eV}$) and, in certain circumstances, beyond. The average DM energy density at one solar distance from the center of our Galaxy is $\rho_{DM} \approx 0.4\, \mathrm{GeV}/\mathrm{cm}^3$ \cite{Olive:2016xmw}, while the energy density in the solar system itself is much less constrained: $\rho_{DM} < 10^5\, \mathrm{GeV}/\mathrm{cm}^3$ \cite{Pitjev:2013sfa}. For the Milky Way, it is assumed that the DM objects form a non-rotating isothermal spherical halo and obey Maxwell distribution over velocities. For the observer in the solar system, the typical velocities of these objects are believed to be $\vv \sim 270$ km/s or $\vv \sim 10^{-3}$ in natural units~\cite{Gelmini:2000dm}, with dispersion $\delta v \approx \vv$. However, since the details of the DM distribution within the solar system are not known, this figure can be different, e.g., comparable to the planetary velocities.

Since the local density of DM is small and, hence, the gravitational effects are too weak to be detected directly, the current methods rely on hypothetic interactions beyond the Standard Model (SM) with the coupling constants that can be varied by many orders of magnitude and still being consistent with the astrophysical, cosmological and gravitational tests. It is constructive to split the aforementioned mass range in two categories: sub-eV and the rest. The latter is usually tested with high-energy and scattering experiments~\cite{Olive:2016xmw}, where typical tested masses of DM particles, the so-called Weakly Interacting Massive Particles (WIMPs), are above 1 GeV. The sub-eV DM, however, due to the large Compton (or de Broglie) wavelength together with given density, is considered as classical fields. Hypothetical interaction of this field with ordinary matter can lead to variation of fundamental constants and hence frequencies of atomic or nuclear transitions, sizes of atoms, etc. Recently, there has been a great interest in the direct detection of such light DM, we refer the Reader to Ref.~\cite{Graham:2015ifn} for a review and to Ref.~\cite{Lee:2017qve} for a brief history of ultralight scalar DM models.

In this paper, we consider the DM in the form of a scalar field. Such choice is motivated partially by the possibility of testing scalar field theories with atomic clocks and partially by the abundance of new light scalar fields predicted by nearly every theory beyond the SM (this is due to the fact that such theories normally undergo a spontaneous symmetry breaking at low energies, which produces massless or nearly-massless bosons). 

While there have been several studies dedicated to the DM tests with atomic clocks, all of them rely on the frequency domain analysis or transient behaviors, hence, can only probe the DM masses smaller than the inverse sampling/interrogation time. We propose a new approach of the time domain analysis of atomic clock stability that is ubiquitously used to characterize the clock performances. An advantage of this approach will be shown to allow us to probe much larger DM field masses (i.e., much faster processes) due to the aliasing of high-frequency perturbations, analogous to the Dick effect~\cite{Dick1987}. As an example of such fast disturbances, we study short-wavelength DM waves and DM topological defects interacting with the clocks. We will also study the sensitivity of the method to the slow processes and compare with existing limits.

The paper is organized in the following way. We begin with the description of the coupling between DM and ordinary matter, classified by an integer $n$, and the effect of this coupling on the atomic clock frequency. We then derive DM signatures in Allan variance plots for various DM objects and interactions. Both existing clock comparisons data and future clock experiments are analyzed and discussed. We mostly focus on the cases $n=1$ and $n=2$, briefly discussing the case of general $n$ when possible.

\section{Atomic clock response}
We describe the DM by a scalar or pseudoscalar field $\phi$ with mass $m_\phi$. We begin with writing down the action (we use the natural units $\hbar=c=1$ here and after),
\begin{align}
S = &\int d^4 x \sqrt{-g} \left\{ \frac{R}{16 \pi G} + \half g_{\mu\nu}\dd^\mu \phi \dd^\nu \phi - V(\phi)\right. \nonumber\\
&~~~~~~~~~~~~~~~~~~~~~~~~~~~~~~~+\left.  \mathcal{L}_{SM}+\mathcal{L}^{(n)}_{int}\right\},
\end{align}
where $G$ is the Newton's constant, $R$ is the Ricci scalar for the spacetime metric $g_{\mu\nu}$, $g \equiv \det g_{\mu\nu}$, $V(\phi) = \half m_\phi^2 \phi^2 + \ldots$ is the scalar field potential, $\mathcal{L}_{SM}$ is the SM Lagrangian and $\mathcal{L}^{(n)}_{int}$ is the interaction Lagrangian, describing the coupling of the $\phi$ field to the usual matter, that we choose in the standard form of a sum over gauge-invariant operators of SM fields coupled to the powers of $\phi$ (see, e.g., Refs.~\cite{Damour:2010rp,Stadnik:2015kia} for particular cases)
\begin{align}
 &\mathcal{L}^{(n)}_{int} = (\pm 1)^{n+1} \phi^n \left[ \frac{1}{4 e^2 \LL{\gamma}{n}^n}  F_{\mu\nu}F^{\mu\nu} - \frac{\beta_{YM}}{2g_{YM} \LL{g}{n}^n}\right.\nonumber\\
 &\times G_{\mu\nu}G^{\mu\nu}\left.-\sum_{f=e, u, d}\left(\frac{1}{\LL{f}{n}^n}+\frac{\gamma_{m_f}}{\LL{g}{n}^n}\right) m_f \bar \psi_f \psi_f\right],\label{Lagrangian}
\end{align}
where $F_{\mu\nu}$ is the standard Maxwell tensor, $G_{\mu\nu}$ is the gluonic field strength tensor, $\beta_{YM}$ is the beta-function for the running of the SU(3) gauge coupling $g_{YM}$, $\gamma_{m_f}$ is the anomalous dimension of the mass operator $\bar \psi_f \psi_f$ for the SM fermions $\psi_f$ (for our energy scales we are considering the electron, u- and d-quarks only, see also comments in Ref.~\cite{Damour:2010rp}). 
We make the $\pm$ sign disappear for odd $n$, because it can be absorbed in the definition of $\phi$ otherwise. Parameters $\LL{a}{n}$ are unknowns of dimension of mass describing the strength of the coupling between the scalar field $\phi$ and SM fields. If Eq.~(\ref{Lagrangian}) is considered as an effective Lagrangian, then $\LL{a}{n}$ describe energy scales for the physics beyond the SM. Parameter $n$ is an integer and $\LL{a}{n}^n$ means $\LL{a}{n}$ raised to power $n$. The case $n=1$ is usually considered in the dilaton DM studies~\cite{Damour:2010rp, Arvanitaki:2014faa, Hees:2016gop}, while $n=2$ is usually considered for the axion or topological DM~\cite{Derevianko:2013oaa, Stadnik:2015kia}. The latter case is suitable for both scalar and pseudoscalar fields, since the field contributes in a parity-even combination $\phi^2$. It is also worth mentioning that the values of the scales $\LL{a}{n}$ become less constrained with the growth of $n$, since they contribute in the power $n$ and the interaction terms become more suppressed.

The Lagrangian (\ref{Lagrangian}) is chosen in the most general form that preserves gauge symmetry and contains a minimal number of new parameters~\footnote{Even though it seems to be an obvious extension, we do not consider the anomalous couplings, $\phi \epsilon^{\mu\nu\alpha\beta}F_{\mu\nu}F_{\alpha\beta}$, $\phi \epsilon^{\mu\nu\alpha\beta}G^a_{\mu\nu} G^a_{\alpha\beta}$ and $(\partial_\mu \phi) \bar \psi \gamma^\mu \gamma^5 \psi$, since they are well constrained by the standard axion searches~\cite{Olive:2016xmw}. In addition, it is not clear if it is possible to test those couplings with atomic clocks.}. 
For other possible DM candidates and couplings (so-called ``portals''), as well as for a general theoretical review, see Ref.~\cite{Graham:2015ifn}. In particular, the linear Higgs coupling $g_h \phi H^\dagger H$, where $H$ is the Higgs doublet, will lead to $\LL{f}{1}=m_h^2/g_h$, $\LL{\gamma}{1} = 8 \pi m_h^2 / (\alpha g_h)$~\cite{Piazza:2010ye}, $\LL{g}{1} = 9 m_h^2/(2 g_h)$~\cite{Graham:2015ifn}, where $m_h=125 $ GeV is the Higgs mass and $\alpha$ is the fine structure constant. The quadratic coupling $g_h' \phi^2 H^\dagger H$ has a structure of the mass-term for $\phi$ and will renormalize the mass $m_\phi$ to the values proportional to the utraviolet-cutoff, so $g_h'$ is required to be (untestably) small in this case~\cite{Piazza:2010ye}.

Coupling of $\phi$ to SM fields, Eq.~(\ref{Lagrangian}), induces changes in the values of various fundamental constants, such as~\cite{Damour:2010rp}
\begin{align}
 &\frac{\delta \alpha}{\alpha} = (\pm 1)^{n+1} \left(\frac{\phi}{\LL{\gamma}{n}}\right)^n, \frac{\delta m_f}{m_f} = (\pm 1)^{n+1} \left(\frac{\phi}{\LL{f}{n}}\right)^n,\nonumber\\
&~~~~~~~~~~~~~~~~\frac{\delta \Lambda_{QCD}}{\Lambda_{QCD}} = (\pm 1)^{n+1} \left(\frac{\phi}{\LL{g}{n}}\right)^n,\label{variations}
\end{align}
where $\alpha$ is the fine-structure constant and $\Lambda_{QCD}$ is the QCD scale.
It is known that an atomic clock response to the variations of the constants can be expressed as \cite{Flambaum:2004tm}
\begin{align}
\nu = \mathrm{const}\cdot R_\infty\cdot \alpha^{K_\alpha}\left(\frac{m_q}{\Lambda_{QCD}}\right)^{K_{q\Lambda}}\left(\frac{m_e}{\Lambda_{QCD}}\right)^{K_{e\Lambda}},\label{response}
\end{align}
where $\nu$ is the clock frequency, $R_\infty$ is the Rydberg constant (or any other atomic energy/frequency unit), 
$m_q = (m_u+m_d)/2$ and exponents $K_a$ are either known or can be expressed through other analogous constants for existing atomic clocks, see Refs.~\cite{Flambaum:2004tm, Flambaum:2008kr, Prestage:1995zz, Arvanitaki:2014faa} and references therein~\footnote{We drop the dependence on the strange-quark mass as it is usually negligibly small for most of the atomic clock transitions.}. For the microwave clocks all three exponents contribute due to the dependence of the hyperfine structure constant on the proton and electron masses, in particular, $K_{e\Lambda}=1$. For the most common frequency standard, ${}^{133}$Cs, $K_{\alpha}=2.83$, $K_{q\Lambda}=0.07$. For optical clocks, $K_{q\Lambda}=K_{e\Lambda}=0$ and $K_{\alpha}\neq 0$, see Table~\ref{opttable} for the values of $K_{\alpha}$. By varying (\ref{response}) and using (\ref{variations}), we obtain
\begin{align}
\ddd\frac{\delta \nu}{\nu} = (\pm 1)^{n+1} \phi^n \left[ \frac{K_\alpha}{\LL{\gamma}{n}^n}+\frac{K_{q\Lambda}}{\LL{q}{n}^n}-\frac{K_{q\Lambda}+K_{e\Lambda}}{\LL{g}{n}^n}\right],\label{freqshift}
\end{align}
where, for the simplicity of the notation,
\begin{align}
\LL{q}{n}^n \equiv \frac{\LL{u}{n}^n \LL{d}{n}^n (m_u + m_d)}{m_u \LL{d}{n}^n + m_d \LL{u}{n}^n}.
\end{align}

\begin{table*}[t]
\centering
\caption{Examples of exponents of the fine-structure dependence (\ref{response}) and short-term stabilities $\sigma_0$ for the clocks mentioned in this article. See more in Refs.~\cite{Flambaum:2008kr, HGclock, HGclock2, NISTAlHg, Bloom:2013uoa, Arvanitaki:2014faa, Campbell:2012zzb, Ybclock, ThK}.}\label{opttable}
\medskip
\begin{tabular}{l | c c c c c c c c c}
\hline
Species & ${}^{133}\mathrm{Cs}$ &${}^{199}\mathrm{Hg}^+$ & ${}^{199}\mathrm{Hg}$ & ${}^{27}\mathrm{Al}^+$ & ${}^{87}\mathrm{Sr}$& ${}^{171}\mathrm{Yb}$& ${}^{162}\mathrm{Dy}$& ${}^{164}\mathrm{Dy}$& ${}^{229}\mathrm{Th}$\\
\hline
States & hyperfine& $\mathrm{5d^96s^2\, {}^2D_\frac{5}{2}}$& $\mathrm{6s6p\, {}^3P_0}$& $\mathrm{3s3p\, {}^3P_0}$& $\mathrm{5s5p\, {}^3P_0}$& $\mathrm{6s^2\, {^1S_0}}$& $\mathrm{4f^95d^26s}$ & $\mathrm{4f^{10} 5d6s}$& nuclear\\
 & hyperfine& $\mathrm{5d^{10}6s\, {}^2S_\frac{1}{2}}$& $\mathrm{6s^2\, {}^1S_0}$& $\mathrm{3s^2\, {}^1S_0}$& $\mathrm{5s^2\, {}^1S_0}$& $\mathrm{6s6p\, {^3P_0}}$& $\mathrm{4f^{10}5d 6s}$ & $\mathrm{4f^{9} 5d^2 6s}$& nuclear\\

\hline
$K_\alpha$ & 2.83 & -3.19 & 0.81 & 0.008 & 0.06 & 0.32 & $8.5\times 10^6$ & $-2.6\times 10^6$ & $10^4 (?)$\\
\hline
$\sigma_0 (10^{-16}\mathrm{Hz}^{-1/2})$& $10^3$ & 28 & 1.8 & 28 & 3.1 & 0.6 & $4\times10^7$& $1\times10^8$ & $10 (?)$\\
\hline
\end{tabular}
\end{table*}

\begin{figure*}[t]
\centering
\subfigure[\label{exclusions1}]{\includegraphics[width=9cm]{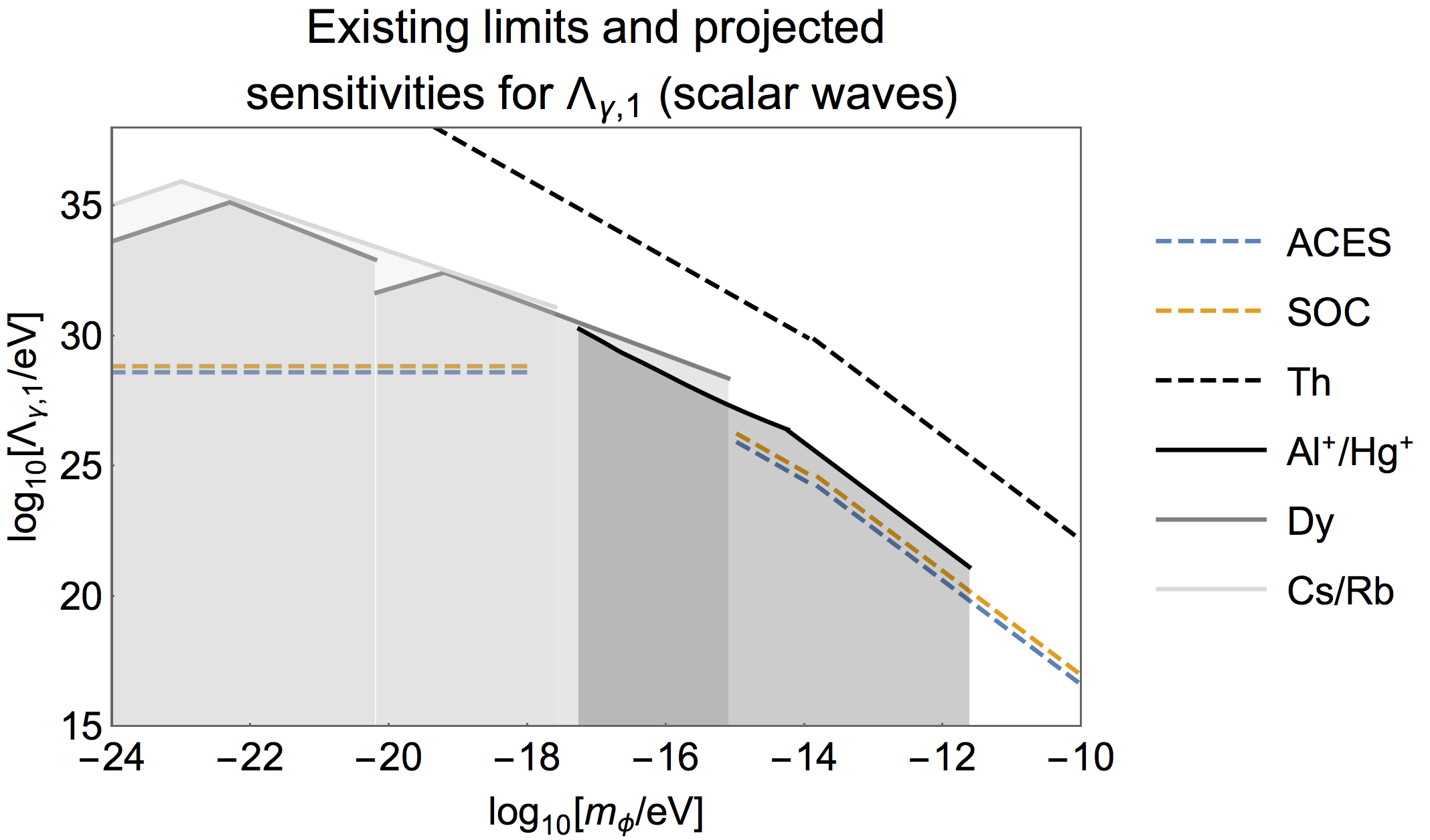}}\hspace{1cm}
\subfigure[\label{exclusions2}]{\includegraphics[width=7cm]{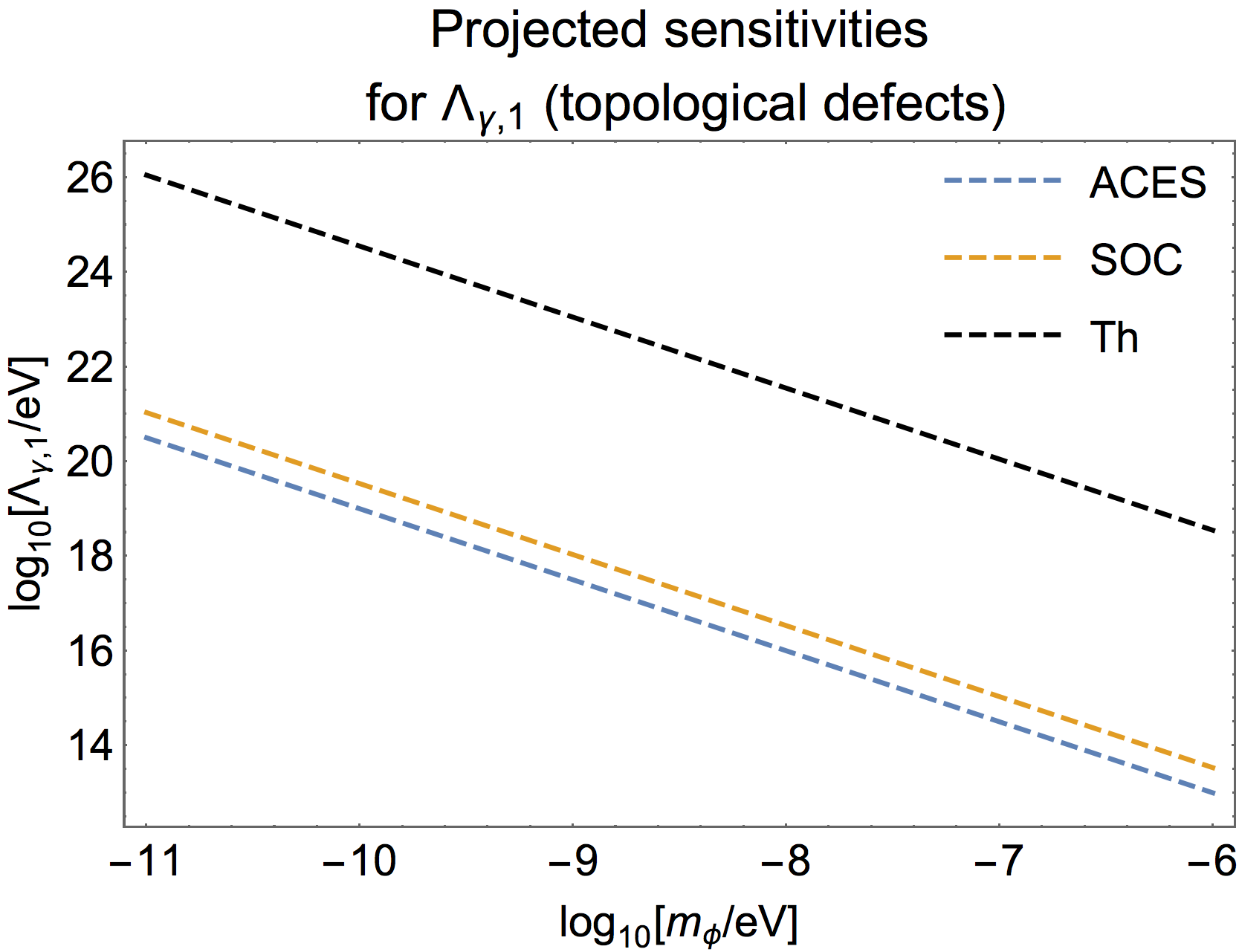}}
\caption{Existing exclusion regions and projected sensitivities in the case of linear coupling, $n=1$. (a) Limits for the DM waves. Solid lines correspond to the existing limits~\cite{VanTilburg:2015oza, Hees:2016gop} (the black line was obtained in this paper), dashed lines correspond to the sensitivities of future and potential experiments; (b) Limits for the topological (monopole/string) DM. It is assumed that $\LL{g}{1}, \LL{q}{1} \to \infty$ for the microwave clocks. \label{exclusions}}
\end{figure*}

\begin{figure*}[t]
\centering
\subfigure[\label{exclusions3}]{\includegraphics[width=8.2cm]{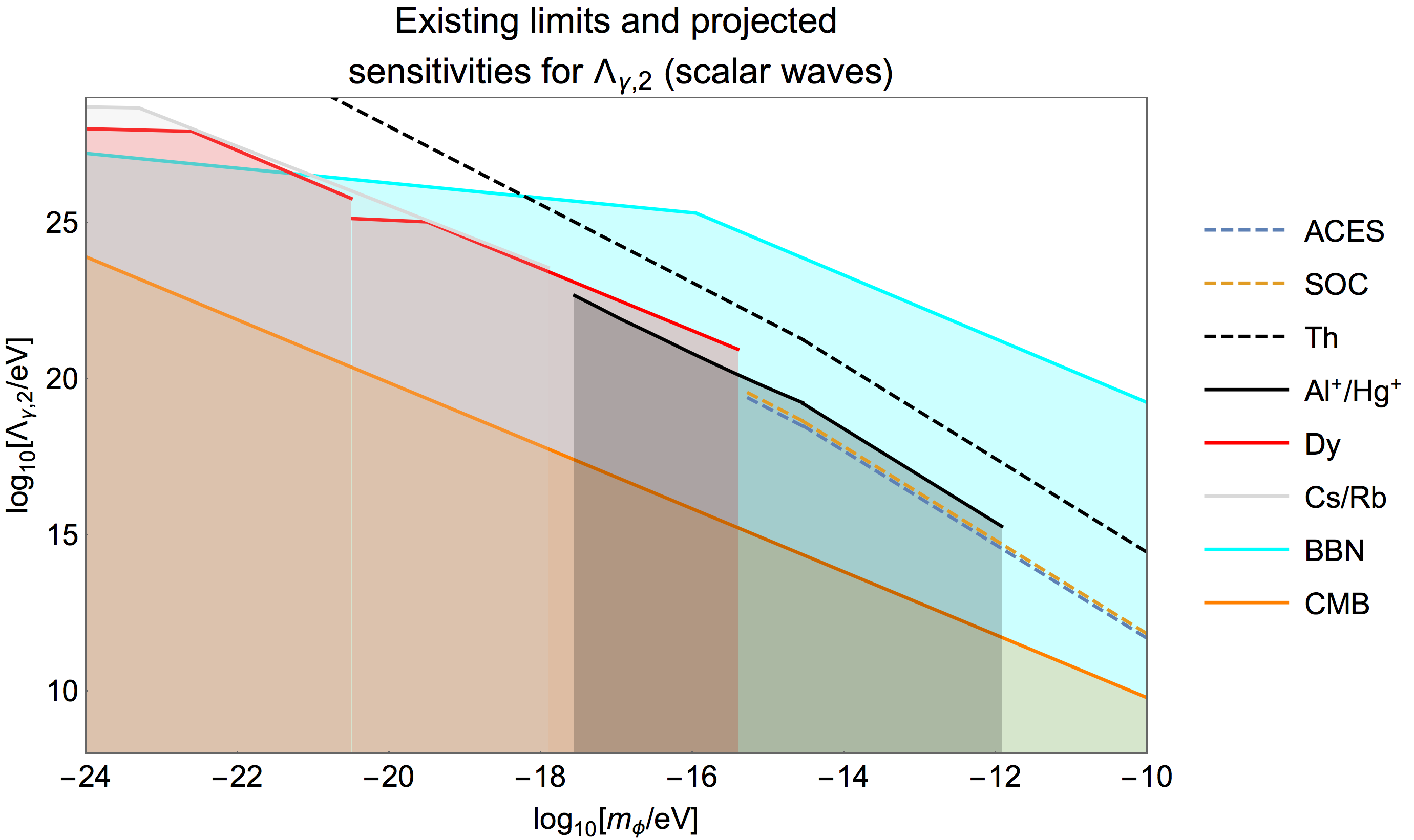}}\hspace{1cm}
\subfigure[\label{exclusions4}]{\includegraphics[width=7cm]{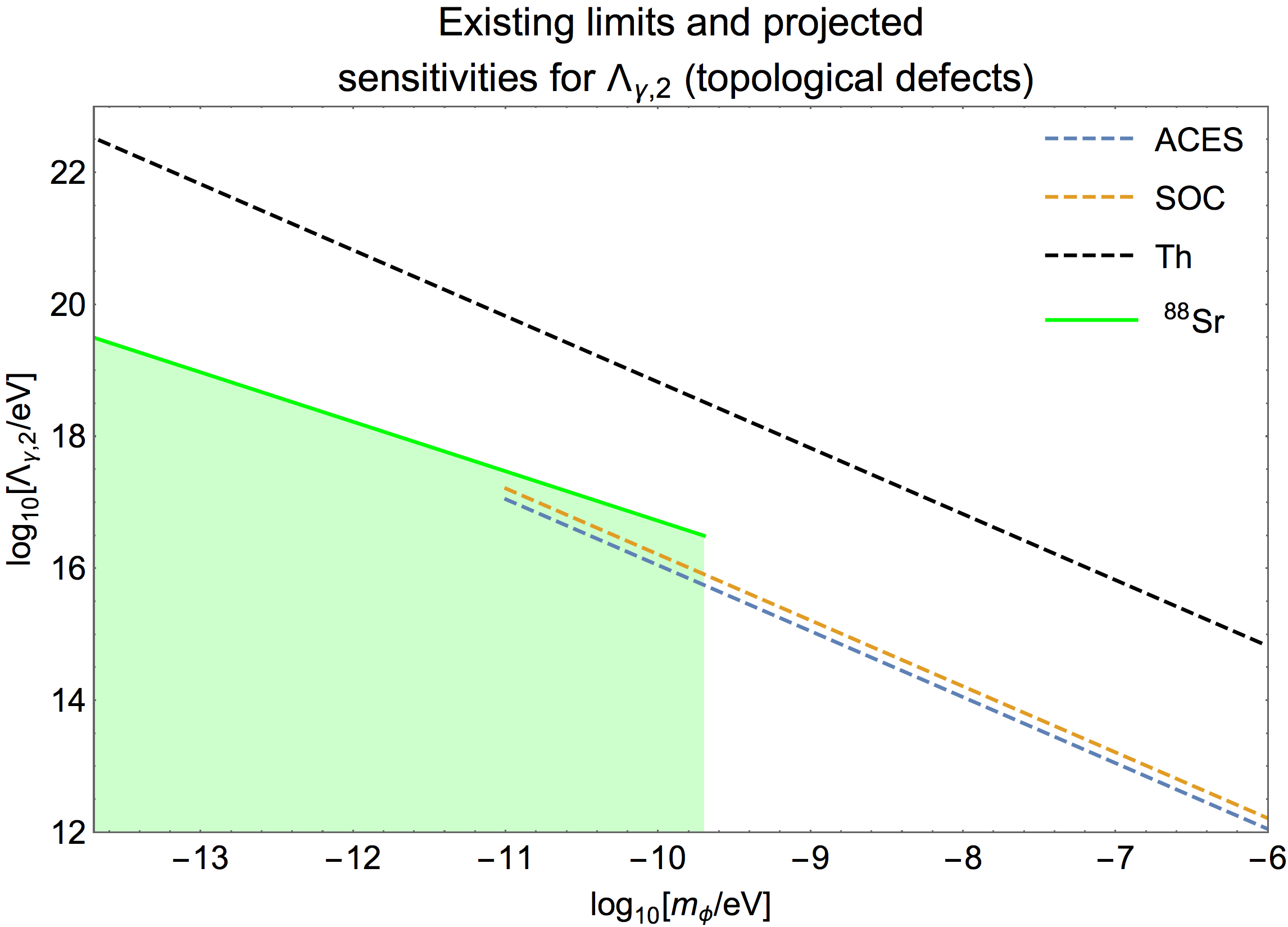}}
\caption{Existing exclusion regions and projected sensitivities in the case of quadratic coupling, $n=2$. (a) Limits for the DM waves. Solid lines correspond to the existing limits (the black line was obtained in this paper, the other lines correspond to the Cs/Rb, Dy, CMB and BBN limits, see Refs.~\cite{Hees:2016gop, Stadnik:2015kia}), dashed lines correspond to the sensitivities of future and potential experiments; (b) Limits for the topological (monopole/string) DM with $\TT=$1 yr. The green region is reproduced from Ref.~\cite{Wcislo:2016qng}. It is assumed that $\LL{g}{2}, \LL{q}{2} \to \infty$ for the microwave clocks. \label{quadexclusions}}
\end{figure*}

Currently, there exist a number of limits on the interaction parameters. Limits on $(m_\phi,\LL{a}{1})$ are given in, e.g., Refs.~\cite{Hees:2016gop, Arvanitaki:2015iga, Leefer:2016xfu} based on various scalar field distributions. 
To interpret some of the figures existing in the literature, one should take into account the following difference in notations: $d_a = \frac{M_{\mathrm{P}}}{\sqrt{4\pi}\LL{a}{1}}$, where $M_{\mathrm{P}}= 1.2\times 10^{19}$~GeV is the Planck mass. 
For the $(m_\phi,\LL{a}{2})$ exclusion diagrams, see Ref.~\cite{Stadnik:2015kia, Wcislo:2016qng}.  For the sake of simplicity, we mostly focus on $\LL{\gamma}{n}$ in this article, and reproduce the relevant limits in Figs.~\ref{exclusions} \& \ref{quadexclusions}. These limits are based on the clock comparison experiments, as well as on the studies of the big-bang nucleosynthesis (BBN) and cosmic microwave background (CMB).

It should be emphasized that the exclusion regions should be compared with a caution - different assumptions on the spatial distribution of DM can lead to different sensitivities, while having the same local average energy density. Here we briefly list some types of DM distribution that can be measured with atomic clocks.

(a) {\it Dark matter waves} usually appear as solutions to the linearized field equations and can be studied with the use of co-located clocks of different type (direct frequency comparison~\cite{Arvanitaki:2014faa, Hees:2016gop, VanTilburg:2015oza}), as well as spatially separated clocks (e.g., via a two-point correlation function~\cite{Derevianko:2016vpm}). For co-located clocks  one exploits different atom transitions (with different $K_a$) so there is a non-vanishing difference in the fractional frequency change between two clocks. Such difference, if induced by DM, will be periodic in time. The wavevector $k = m_\phi \vv = 10^{-3} m_\phi$ is small enough to neglect the spatial variation of the field on the scale of experiment. The range of masses is usually bounded from above by the inverse atomic clock loop time, at which the clocks are referenced to the atomic transition, usually, $\tau_{loop} \sim  1$ s for high-performance clocks, so $m_\phi < 2 \pi / \tau_{loop} \sim 10^{-14}$ eV. One can circumvent this limit by comparing a high short-term stability atomic oscillator (such as Ca clock~\cite{CaNIST}) to a mechanical oscillator (e.g., a stabilized Fabry-P\'{e}rot cavity). Resonant frequencies of the mechanical oscillators also depend on the presence of DM, since the atomic unit of length, $a_{Bohr} = \hbar (\alpha m_e c)^{-1}$, and, therefore, sizes of atoms and solid bodies vary together with the fundamental constants~\cite{Wcislo:2016qng, Arvanitaki:2015iga}. One can also study modification of the gravitational potential by the DM waves~\cite{Khmelnitsky:2013lxt, Porayko:2014rfa}, however, this type of measurement belongs to the domain of the gravitational wave studies and is beyond our scope. 

(b) {\it Clumps of dark matter}, in particular, topological defects (0D monopoles, 1D strings, 2D domain walls) that can result from, e.g., phase transitions in the early universe. Usually, such objects are classical solutions of the field equations with the space of vacua (minima of potential energy density) characterized by a nontrivial homotopy group (e.g., for a string, $\pi_1(S^1) = \mathbb{Z}$). Field $\phi$ itself can be a real or complex field, or a field multiplet, see Ref.~\cite{Rubakov:2002fi} for examples. Such defects moving with galactic velocities can be studied by a transcontinental (or space) network of atomic clocks~\cite{Derevianko:2013oaa, Roberts:2017hla}.
It is proposed to track consecutive intermittent phase changes for atomic clocks separated by a large distance, while the DM clump moves through the network. Such approach has an advantage of measuring the time the DM objects sweep through the network, instead of comparing spike-like frequency changes in the frequencies of co-located clocks.
The sensitivity for a system of two identical optical clocks experiencing a consecutive interaction with a DM object can be estimated by (SNR = 1) \cite{Derevianko:2013oaa}
\begin{align}
\LL{\gamma}{2} >  \frac{d\, l^{1/4} (\rho_{DM}\,\TT K_\alpha)^{1/2}}{(2 \vv)^{1/4}\tau^{3/4} \sigma_y^{1/2}(\tau)} \propto \frac{l^{1/4} K_\alpha^{1/2}}{\tau^{3/4} \sigma_y(\tau)^{1/2}} ,\label{sensitivity}
\end{align}
where $d\sim \hbar/(m_\phi c)$ is the size of the topological defect, $l$ is the distance between clocks, $\tau$ is the period of comparison measurements, $\sigma_y(\tau)$ is the Allan deviation $\TT$ is an average time interval between close encounters with the DM objects. As an example, the sensitivity curve for a system of two ${}^{87}$Sr clocks, placed on different continents with $\TT = 1$ yr \cite{Derevianko:2013oaa}, will be right on the upper edge of the BBN exclusion region. 
In the r.h.s. of Eq.~(\ref{sensitivity}) we also factored out the parameters that can be controlled in an experiment. It immediately brings us to the conclusion that high $K_\alpha$, large distance between clocks and better clock stability lead to better sensitivities and higher limits on the scale $\LL{\gamma}{2}$. For instance, transition from Sr to Dy with increase in $K_\alpha$ by 8 orders of magnitude does not lead to a significant improvement of the sensitivity due to the current poor short-term stability of Dy clocks, see Table~\ref{opttable}. However, with the improvement of technology and, hence, the stability in the future, the use of Dy would be highly beneficial.  
In this paper we propose an alternative method for spatially separated clocks that is independent of distance.

(c) One can study {\it caustics} of DM created by the microlensing (or focusing) of DM streams by massive bodies~\cite{Prezeau:2015lxa}. Such focusing can amplify the DM energy density by many orders of magnitude. For Earth it will be a factor of $10^7$ with the beginning (``root'') of the caustic being located around $10^6$ km from Earth. The thickness of the caustic depends on the velocity dispersion for the DM momentum distribution. If the solar system has its own halo of DM, the root might be much closer to the Earth center (the distance to the root is proportional to the square of DM stream velocity). An atomic clock sent to such caustic could experience sudden change in frequency by many orders of magnitude, comparing to the non-caustic measurement. However, uncertainty in the position of the root (due to the uncertainty in the local DM velocity and direction of motion) makes such mission not very optimistic.

(d) {\it Static distribution} around celestial bodies. One can study this case with fifth-force type experiments and search for periodic changes in the frequencies of atomic clocks due to the change in the distance between the clocks and the source of DM, see, e.g., Ref.~\cite{Leefer:2016xfu}. One of the examples would be ground-based searches for seasonal variations of the frequencies due to the changes in the distance between Earth and Sun. 

(e) Finally, one can consider a {\it stochastic background} of DM waves and measure it by means of a network of precision measurement tools. The first method (monochromatic DM wave detection) assumed the entire energy density of DM being carried by a monochromatic wave with a frequency fixed at $m_\phi/2\pi$. However, if the total energy density is distributed over a range of frequencies, then the limits presented in Fig.~\ref{exclusions1} will be significantly reduced. Therefore, it makes sense to put limits not only on the DM couplings, but also on the spectrum of DM excitations~\cite{kalaydzhyan}.

\section{Anomalies in the Allan deviation data}

From previous discussions, it is clear there are various kinds of DM field configuration and various atomic responses. A specific design of experiments and data analysis can be tailored for optimization, either in the time domain or in the frequency domain. In this section, we analyze the possible DM signatures in Allan variance plots. In some cases where the DM signatures are of high frequency, the stability analysis proves to be a more robust detection scheme. Results of this section can be applied to any new scalar field, not just scalar DM. 

The experiments are performed by phase and frequency comparisons between two clocks. In the case of DM waves, one of the clocks is assumed to have a much smaller $K_a$ than the other (for simplicity; otherwise one has to consider the difference between the two $K_a$). In the case of topological defects, such ``reference'' clock can have any $K_a$, but the separation between the two clocks should be much larger than the DM inhomogeneity scale $1/m_\phi$, so the defect does not interact with the two clocks at the same time.
If the experiment is performed in space, we do not consider noises related to the signal transfer due to the refractive index fluctuations (in the ionosphere, troposphere and interplanetary medium) -- we expect our results to be of qualitative rather than quantitative nature. The clock comparison measurement is performed in several sessions, and data from each session are used to produce the stability diagram. This approach is useful if the experiment has long dead times, such as the frequency comparison between ground-based clocks and clocks on low Earth orbit.

\begin{figure*}[t]
\centering
\subfigure[\label{pic1}]{\includegraphics[width=7cm]{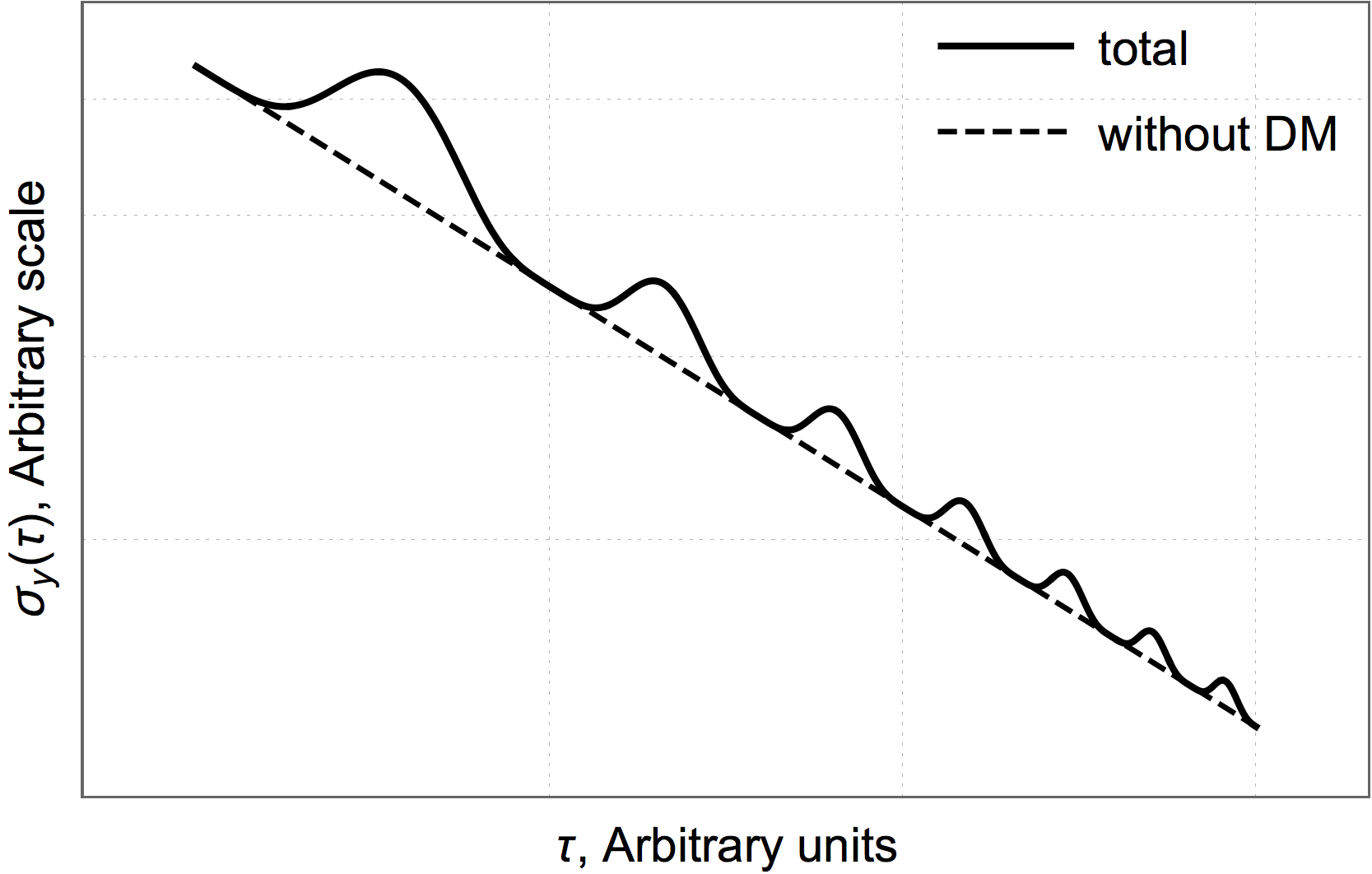}}\hspace{1cm}
\subfigure[\label{pic2}]{\includegraphics[width=7cm]{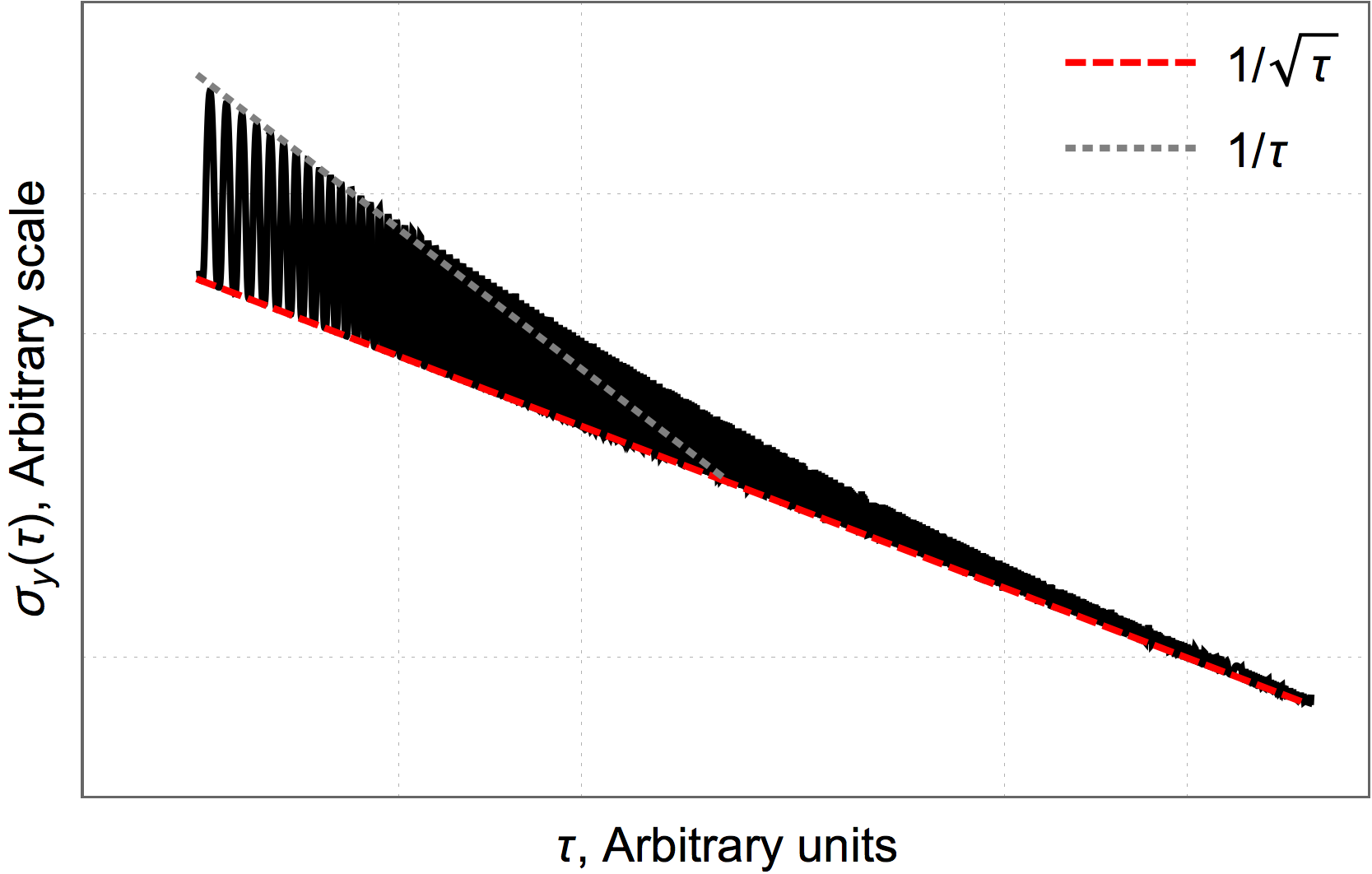}}
\subfigure[\label{pic3}]{\includegraphics[width=7cm]{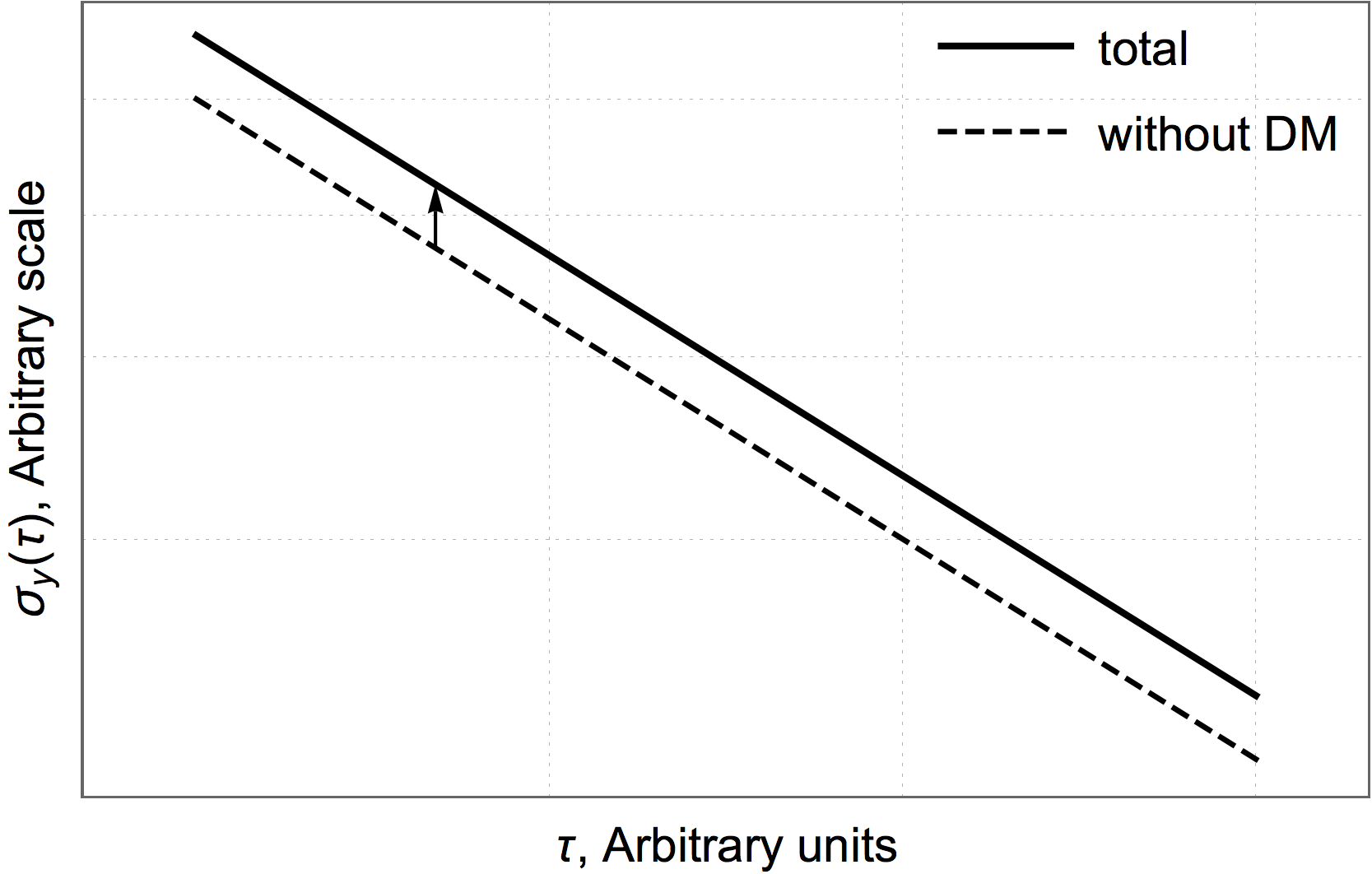}}\hspace{1cm}
\subfigure[\label{pic4}]{\includegraphics[width=7cm]{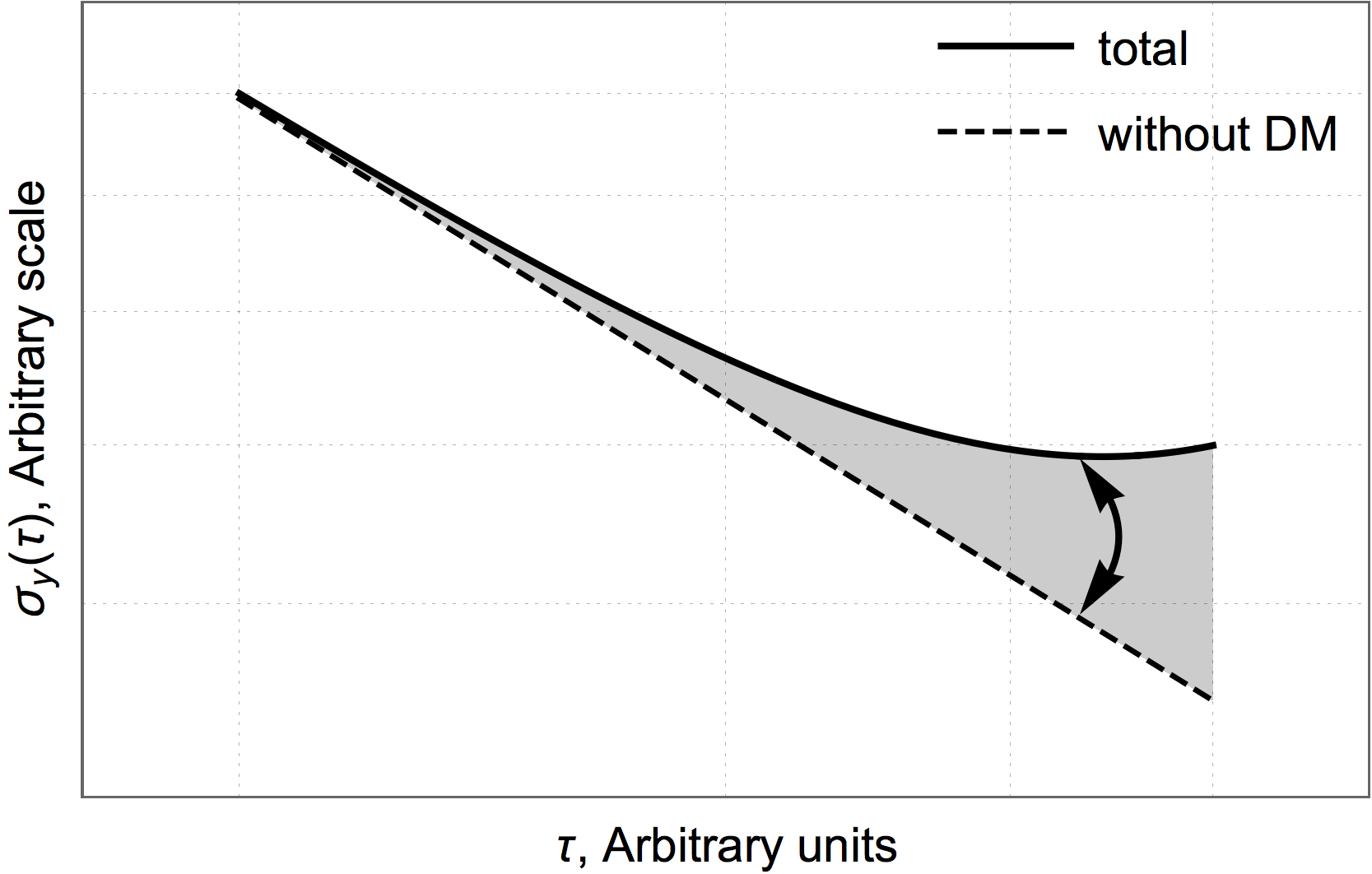}}

\caption{Anomalies in the Allan deviation curve (so-called, $\sigma-\tau$ diagram): (a) Bumps due to the periodic frequency variation (modulating frequency, $f$, comparable to the inverse averaging time, $f \in [1/\tau_{max}, 1/\tau_{loop}]$); (b) Deviation from the expected slope (a train) at short averaging times -- $f \gg 1/\tau_{loop}$; (c) Shift of the curve up due to an additional white noise or a single phase jump; (d) Additional linear (or quadratic) deviation changing periodically in time due to slow frequency variation (characteristic time much larger than a single comparison measurement time).\label{anomalies}}
\end{figure*}

The fractional frequency deviation between two clocks is given by $y(t) = d x(t) / d t$, where $x(t)$ is the relative time deviation between two clocks' readings, $x(t) = \frac{\varphi_1(t)}{2 \pi \nu_1} - \frac{\varphi_2(t)}{2 \pi \nu_2}$ and $\varphi_i(t)$ and $\nu_i$ being the phase and the frequency of the clocks, respectively. The average fractional frequency deviation is a function of time,
\begin{align}
\bar y(t) = \frac{1}{\tau}\int\limits_{t - \tau}^t y(t')\,d t'\,,
\end{align}
and defines the {\it Allan variance}~\cite{Allan0}, $\sigma_y^2(\tau)$,
\begin{align}
\sigma_y^2(\tau) = \frac{1}{2} \lim\limits_{T \to \infty} \frac{1}{T} \int\limits_0^T \left[ \bar y(t+\tau) -  \bar y(t) \right]^2 dt\,,\label{oldAllan}
\end{align}
where $\tau$ is the averaging time and $T$ is the measurement time.
Square root of the Allan variance is usually called {\it Allan deviation}. For the general properties of the Allan deviation see Ref.~\cite{Allan}. In what follows, for practical reasons, we replace the limit by the condition $T = p \tau \gg \tau$, $p \in \mathbb{N}$ and consider the Allan variance as a function of time, at which it was calculated,
\begin{align}
\sigma_y^2(t, \tau) = \frac{1}{2 T} \int\limits_t^{t+T} \left[ \bar y(t'+\tau) -  \bar y(t') \right]^2 dt'\,,\label{Allandef}
\end{align}
which is similar, in some sense, to the dynamic Allan variance~\cite{DAVAR}. Large value of $p$ (and, hence, $T \gg \tau$) is needed for $\sigma_y(\tau)$ to have a small statistical error.
Periodic variation in one of the clock's readings leads to $y(t) = A \cos (2\pi f t + \varphi_0)$ and the average fractional frequency deviation
\begin{align}
\bar y (t) = \frac{A}{\pi f \tau} \sin(\pi f \tau) \cos(\pi f (2t - \tau) + \varphi_0)\,.
\end{align}
Substituting this to Eq.~(\ref{Allandef}) and applying trigonometric identities, we get
\begin{align}
\sigma_y^2(t,\tau) = \frac{2 A^2 \sin^4(\pi f \tau)}{(\pi f \tau)^2}\cdot \frac{1}{T}\int\limits_t^{t+T}\sin^2(2\pi f t' + \varphi_0)\,d t'\,.\label{timeAllan}
\end{align}

In the following sections, we consider separately three situations defined by the hierarchy of three time scales: the continuous measurement time $T$ for each comparison session, the averaging time $\tau$, and the DM interaction timescale $1/m_\phi$. It is assumed that the clock is DM-sensitive only when referenced to an atomic transition frequency. Furthermore, we require $\tau \in [\tau_{loop}, \tau_{max}]$, with $\tau_{max} \ll T$, as sufficient Allan variance statistics is needed to extract meaningful measurement data.

\subsection*{Case $\tau_{loop} < 1/m_\phi < \tau_{max}$}

\begin{figure*}[t]
\centering
\subfigure[\label{bumps}]{\includegraphics[width=7cm]{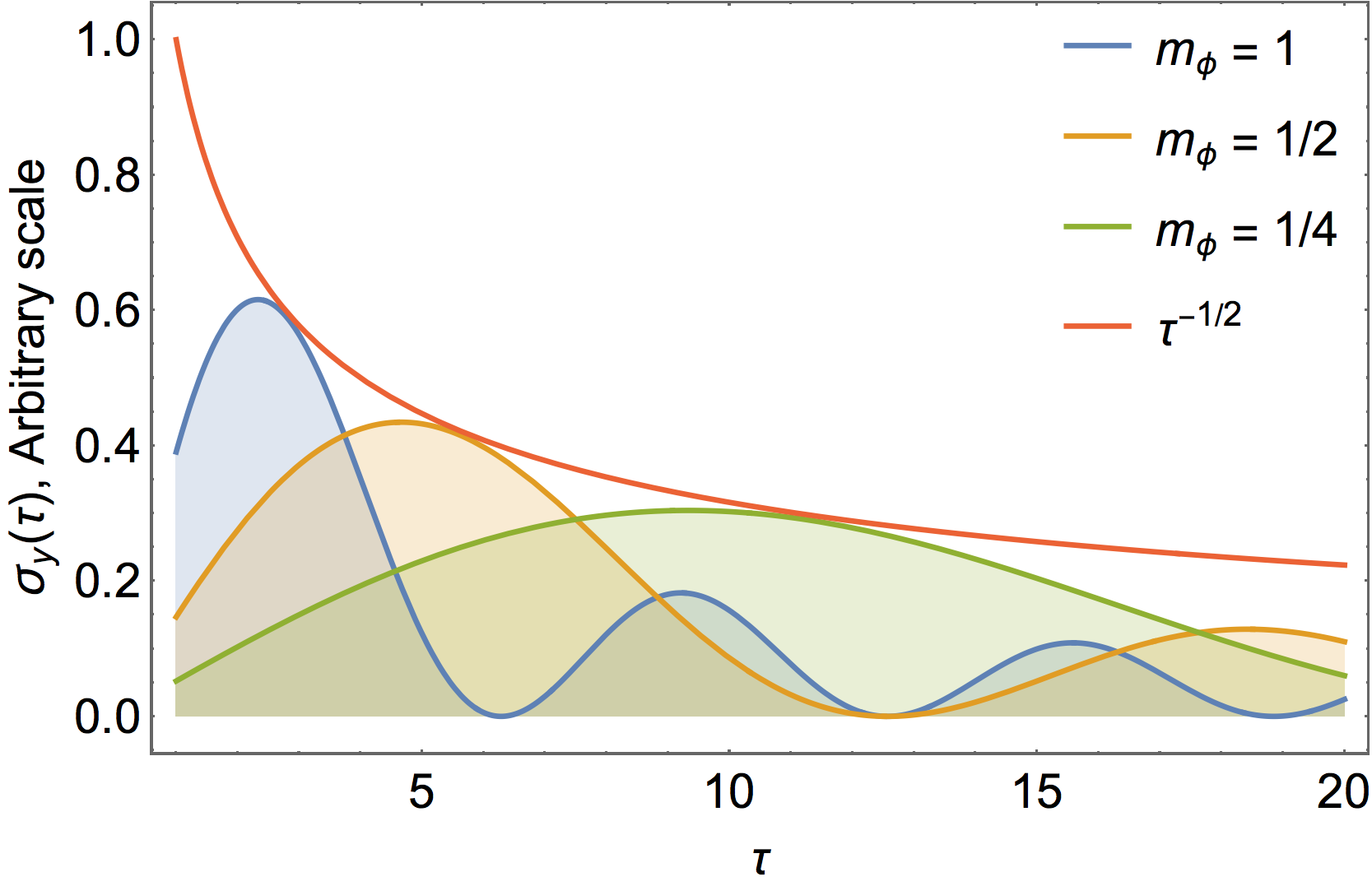}}\hspace{5mm}
\subfigure[\label{slope}]{\includegraphics[width=7cm]{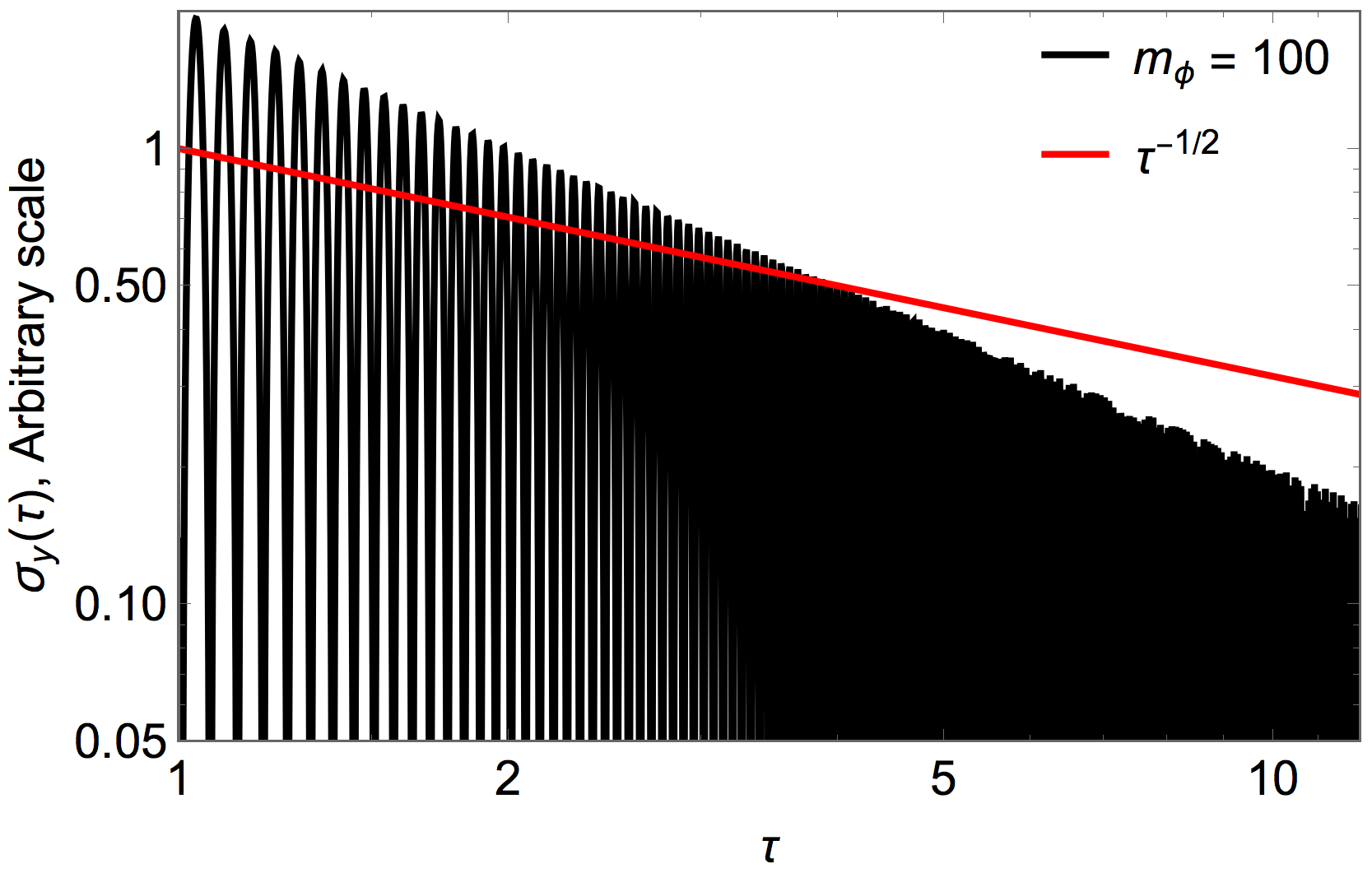}}
\caption{(a) Functions $\sigma_y(\tau) \propto \sin^2(m_\phi \tau/2)/(m_\phi^2 \tau/2)$ scaled to fit under $f(\tau) = \tau^{-1/2}$; (b) Same function for $m_\phi = 100$ in log-log-scale (black) together with $f(\tau) = \tau^{-1/2}$ (red).}
\end{figure*}

In this section, we assume that the DM exists in a form of waves with frequency $f=m_\phi/(2\pi)$ and wavenumber $k = \vv m_\phi \approx 10^{-3} m_\phi$. Since one of the clocks is insensitive to DM, the spatial profile of the wave, and hence $k$, play no role, and we neglect it in the derivation. 
The power spectrum density for the field can be written as $S_\phi(f)=S_\phi' \delta (f - m_\phi/(2\pi))$. Here $S_\phi'$ can be estimated from the DM average energy density $\rho_{DM}\approx (0.04\, \mathrm{eV})^4$,
\begin{align}
\rho_{DM} \sim m_\phi^2 \langle \phi^2 \rangle = m_\phi^2 \int\limits_0^\infty S_\phi (f)\, df = m_\phi^2 S_\phi'\,,
\end{align}
so $S_\phi(f) = \frac{\rho_{DM}}{m_\phi^2}\, \delta (f - m_\phi/(2\pi))$. In order to find a DM-induced Allan variance, we can apply Eq.~(\ref{timeAllan}) with $A^2 = \KK_{(1)}^2 S_\phi'$ and the integral being reduced to $T/2$ in this limit,
\begin{align}
\sigma^2_y(t, \tau)=\sigma^2_y(\tau) =\KK_{(1)}^2 \frac{\rho_{DM}}{m_\phi^2}\frac{\sin^4(m_\phi \tau/2)}{(m_\phi \tau/2)^2}\,.\label{bumpfunc}
\end{align}
This expression has an immediate consequence - if the clock was sensitive to the scalar DM, it would have a primary bump at $\tau \approx 2.3/m_\phi$ in the $\sigma-\tau$ stability diagram. When this additional clock instability is added to the typical well-behaved atomic clock stability of $1/\sqrt{\tau}$, we get the stability curve as shown in Fig.~\ref{pic1}. 
Since such feature was not observed, we can put limits on the DM coupling from short-term Allan deviation $\sigma_y(\tau) = \sigma_0/\sqrt{\tau}$ for existing clocks, see Table~\ref{opttable} for examples of $\sigma_0$. We will require that the DM-induced deviation is smaller than the deviation observed for a given clock~\footnote{This can be improved by fitting the bumps into the measurement errorbars. However, when applied to the existing data, such approach will put limits only on a finite set of masses, corresponding to the set of used averaging times. Continuous set of $\sigma_y(\tau)$ would be beneficial for the future studies.}, see Fig.~\ref{bumps}. Such requirement is justified by the fact that our goal is to put upper limits on the DM-couplings rather than to claim DM detection. The latter would require a proper statistical inference and hypothesis testing, as well as a detailed discussion of possible systematics specific to the space atomic clocks that can mimic the DM signal. Such discussion is out of the scope of this paper; however, we want to mention that GPS clocks experience such periodic variations that differ from their
orbital period~\cite{GPSvar}. 

Considering optical clocks (or microwave clocks with $\LL{q}{1}, \LL{g}{1} \to \infty$), we substitute $\KK_{(1)}=K_\alpha/\LL{\gamma}{1}$ and obtain
\begin{align}
\LL{\gamma}{1} > \frac{2 |K_\alpha| \rho_{DM}^{1/2}}{\sigma_0 m_\phi^2}\cdot \frac{\sin^2(m_\phi \tau/2)}{\tau^{1/2}}\,.
\end{align}
To get the limits, we choose $\tau \approx 2.8 / m_\phi$ that maximizes the right-hand side of the expression (notice slight difference from $2.3/m_\phi$ for the position of the bump). After substituting the numbers and performing conversion of units~\footnote{In this paper we often rely on the following unit conversion rules: $1\,\mathrm{s}^{-1}=1\, \mathrm{Hz} = 4.14\times10^{-15} \mathrm{eV}$ and $1\,\mathrm{km}^{-1} = 1.97\times10^{-10} \mathrm{eV}$.} we finally obtain
\begin{align}
\LL{\gamma}{1} > 10^{-10} |K_\alpha| \left[\frac{1\, \mathrm{Hz}^{-1/2}}{\sigma_0} \right]\cdot \left[\frac{1\, \mathrm{eV}}{m_\phi}\right]^{3/2},\label{optlimit}
\end{align}
where $m_\phi$ should be chosen such that $2.8 / m_\phi$ belongs to the allowed range of averaging times for the given clock. 

To demonstrate an application of the method, we choose the ${}^{199}$Hg clock \cite{HGclock, HGclock2} with $K_\alpha = 0.81$~\cite{Flambaum:2008kr} and $\sigma_0 = 4\times 10^{-15} \mathrm{Hz}^{-1/2}$ when compared to ${}^{87}$Sr clock ($K_\alpha = 0.06$). 
Applying Eq.~(\ref{optlimit}) at $m_\phi = 6\times 10^{-18}$ eV (i.e., $\tau = 2\times 10^3$ s), we obtain $\LL{\gamma}{1} > 10^{30}\, \mathrm{eV}$, which agrees with the current limit obtained from Dy spectroscopy, see Ref.~\cite{VanTilburg:2015oza} (for smaller masses, one has to compare with Ref.~\cite{Hees:2016gop}). It can be further improved by comparing ${}^{27}\mathrm{Al}^+$ ($K_\alpha = 0.008$) ion clock with ${}^{199}\mathrm{Hg}^+$ ($K_\alpha = -3.19$) clock at the National Institute of Standards and Technology, see Ref.~\cite{NISTAlHg} for the description of the experiment and the data. Taking $\sigma_0 = 3.9\times 10^{-15} \mathrm{Hz}^{-1/2}$, we obtain $\LL{\gamma}{1} > 6\times 10^{30}$ eV for the same mass, which can be translated to
$d_e < 6\times 10^{-4}$ in notations of Ref.~\cite{VanTilburg:2015oza}, which is comparable to their limit. In the region of larger masses (up to $10^{-14}$ eV), however, one can establish new limits from the same clocks, see Fig.~\ref{exclusions1}. This combination of two clocks turns out to be ideal for the DM studies due to the precision and almost three orders of magnitude difference in $K_\alpha$. Our method is valid not only for a single monochromatic and coherent wave but also for an isotropic background of waves of the same frequency. The isotropic property is necessary for $S_\phi(f)$ being independent of $\hat k$.

The ideal case scenario, as seen from Eq.~(\ref{optlimit}), corresponds to a combination of large $K_\alpha$ with a high stability, i.e.,  small $\sigma_0$. Among future candidates is a nuclear ${}^{229}$Th clock~\cite{Campbell:2012zzb} based on an exceptionally low excited state of the thorium nucleus (comparing to the typical nuclear energy scales). With $K_\alpha \sim 10^4$~\cite{ThK} and $\sigma_0 \sim 10^{-15}\,\mathrm{Hz}^{-1/2}$ it can surpass the sensitivity of a strontium optical clock by 5 orders of magnitude. Even though such clock is under development at the time this paper was written, we include corresponding projected sensitivities to our plots.

In case of a single coherent monochromatic wave, one can also produce limits for the quadratic coupling, $n=2$. Assume the scalar field at the given point is given by
\begin{align}
\phi(t) = \Phi \cos(m_\phi t + \varphi_0)\,,
\end{align}
then one can exploit the power reduction formula to write
\begin{align}
y(t) = \KK_{(2)} \phi^2(t) = \KK_{(2)} \left[\frac{\Phi^2}{2}+ \frac{\Phi^2}{2} \cos(2m_\phi t + 2\varphi_0)\right]\,.
\end{align}
The first term in the bracket is an unobservable constant shift and can be absorbed in the definition of the clock frequency, while $\pm$ sign in $\KK_{(2)}$ can be absorbed in $\varphi_0$. Noting that $\rho_{DM} = m_\phi^2 \langle \phi^2 \rangle = m_\phi^2 \Phi^2 /2$, we get
\begin{align}
A=\frac{|\KK_{(2)}|\rho_{DM}}{m_\phi^2},\qquad f = m_\phi/\pi\,.
\end{align}
This can be further substituted in Eq.~(\ref{timeAllan}) and yields
\begin{align}
\sigma_y(\tau) = \frac{|\KK_{(2)}|\rho_{DM}\sin^2(m_\phi \tau)}{m_\phi^3 \tau}\,.
\end{align}
Assuming, for simplicity, the sensitivity of the clock to the variations of $\alpha$ only, as well as $\sigma_y(\tau)\propto \tau^{-1/2}$, we repeat the previous analysis to get
\begin{align}
\LL{\gamma}{2}>\frac{|K_\alpha|^{1/2}\rho_{DM}^{1/2}|\sin(m_\phi \tau)|}{\sigma_0^{1/2}m_\phi^{3/2}\tau^{1/4}}\,,
\end{align}
and, therefore (optimal $\tau \approx 1.4 / m_\phi$),
\begin{align}
\LL{\gamma}{2} > 3.7\times 10^{-7} |K_\alpha|^{1/2} \left[\frac{1\, \mathrm{Hz}^{-1/2}}{\sigma_0} \right]^{1/2}\cdot \left[\frac{1\, \mathrm{eV}}{m_\phi}\right]^{5/4}.
\end{align}
We plot the sensitivity curves for ${}^{133}$Cs (future ACES experiment~\cite{Delva:2012zk}, $\sigma_0 = 10^{-13}\, \mathrm{Hz}^{-1/2}$) and ${}^{87}$Sr optical clocks (e.g., future SOC
experiment~\cite{Origlia2016}, $\sigma_0 = 10^{-15}\, \mathrm{Hz}^{-1/2}$) aboard the International Space Station (ISS) together with exclusions in Fig.~\ref{exclusions3} (we also translated $n=1$ limits from existing clock comparison experiments to $n=2$ case). Whenever performing a measurement with ISS atomic clocks, we use the data collected during the ISS flyby only, which puts limitations on the maximal available averaging times and, therefore, minimal testable DM mass for this method. 
It was explicitly assumed that one of the clocks is much less sensitive to DM than the other. One can consider, instead, two identical clocks separated by a distance $D = |\vec D|$. In the case of DM waves with $n=1$, this can be shown to be equivalent to 
\begin{align}
A = \frac{2 \KK_{(1)}\rho_{DM}^{1/2}}{m_\phi}\cdot \sin\left[ \frac{m_\phi \vec v_g \cdot \vec D}{2}\right]\,,\\
f = m_\phi / (2\pi), \quad \varphi_0 = m_\phi \vec v_g \cdot \vec D /2\,,
\end{align}
where we simply considered the difference in the fractional frequency variation between two clocks. The limit $D \to 0$ makes the system insensitive to DM, since the frequency variation happens in phase for both clocks. The best sensitivity is achieved when $\vec D = \frac{\pi q}{\vv^2 m_\phi} \vec v_g$, and $q$ are odd numbers, i.e., when the frequency variation happens in opposite phase. This improves the sensitivity only by a factor 2 comparing to the case with different clocks considered above and, taken the uncertainty in $\hat v_g$, does not seem to be practically advantageous.

\subsection*{Case $1/m_\phi < \tau_{loop}$}

The analysis from the previous section can be applied to the DM fields with mass much larger than the inverse of the clock loop time. In this case one should compare the Alan deviation curve with one of the higher order bumps of function~(\ref{bumpfunc}). While the accuracy of the method discussed in the previous section increases with the growing maximal averaging time, in the case of large $m_\phi$, one should instead look at the small averaging times, where the DM influence on the clock will manifest itself in a train with maxima placed on $\tau^{-1}$ line at small $\tau$,
see Figs.~\ref{pic2} \& \ref{slope}. The absence of such effect gives us
\begin{align}
\LL{\gamma}{1} > \frac{2 |K_\alpha| \rho_{DM}^{1/2}}{\sigma_0 m_\phi^{2}  \tau_{loop}^{1/2}}, \qquad
\LL{\gamma}{2} > \frac{|K_\alpha|^{1/2} \rho_{DM}^{1/2}}{\sigma_0^{1/2} m_\phi^{3/2}  \tau_{loop}^{1/4}}\,.\label{EQslope}
\end{align}
One can show that this limit can be reproduced by fitting secondary bumps under the $\sigma_y(\tau)=\sigma_0/ \sqrt{\tau}$ curve. Projected sensitivities are demonstrated in Figs.~\ref{exclusions1} \& \ref{exclusions3}. We also plot the exclusion region for Al${}^+$/Hg${}^+$ clock comparison obtained from Eq.~(\ref{EQslope}). To our knowledge, this is the first direct limit on the DM waves with linear coupling in this region.

Maximum DM mass (wave frequency) that can be probed with this method is defined by the coherence time for the wave~\footnote{We are grateful to Peter Wolf, who pointed this out to us and showed how the known in the literature result~(\ref{coherence}) can be obtained.}. Since we consider a nonrelativistic massive scalar wave with dispersion relation $\omega = m_\phi + k^2/(2 m_\phi)$, the coherence time $\tau_c$ is given by
\begin{align}
\tau_c = \frac{2\pi}{\delta \omega} = \frac{2\pi}{\omega v_g \delta v}\approx \frac{2\pi \times 10^6}{\omega}\,,\label{coherence}
\end{align}
i.e., the wave is coherent over approximately $10^6$ oscillation periods. For our purposes (e.g., clock comparison during the ISS passage), $\tau_c > 300$ s translates to $m_\phi < 10^{-10}$ eV, which limits $m_\phi$ from above in Figs.~\ref{exclusions1} \& \ref{exclusions3}. Additionally, 11000 s data for the $\mathrm{Hg}^+/\mathrm{Al}^+$ clock comparison, Ref.~\cite{NISTAlHg}, limits the DM masses probed by the experiment to $m_\phi < 2\times10^{-12}$ eV.

One may ask how one can observe the atomic responses to the DM fields at time scales shorter than the clock loop time where the clock frequency is determined by the local oscillator. This can be understood as high frequency aliasing effect. Aliasing effect in atomic clock stabilities is well analyzed by J. Dick~\cite{Dick1987}. Only in our analysis, a sinusoidal disturbance is assumed. In fact, if one of the clocks is subject to a fast oscillating magnetic field, a similar feature will be seen in the instability plot. The coefficients $K_a$ in this regime can be slightly different due to the additional local oscillator response to DM~\cite{Wcislo:2016qng, Roberts:2017hla}. To draw precise conclusions, the effect of the control loop should be analyzed for the clocks in consideration~\cite{Daphna}. Since we are focusing mostly on the order of magnitude estimates, such analysis will be performed elsewhere.

In the high frequency regime,  it also makes sense to study compact DM objects of the typical size of $1/m_\phi$ and average scalar field $\bar \phi$ inside, which interact with one of the clocks during time $t_{int}\sim 1/(m_\phi \vv)\approx 10^3/m_\phi$. Such objects can be, for example, monopoles and strings~\cite{Derevianko:2013oaa} -- stable topological solitons~\footnote{It is important to distinguish the term ``soliton'' used to describe nonperturbative solutions in the gauge field theories from the exact solutions of integrable systems. While in the latter case the shape of the solitons is protected by an infinite number of conservation laws, so they can pass through each other without being distorted, in the former case, the solitons are defined as objects with localized distribution of energy density, and the shape of such objects can change drastically after interacting with each other. This is one of the reasons we do not consider domain walls in this paper.}. We can choose $t_{int} \ll \tau_{loop}$ and treat the effect of clock interaction with such object as a phase jump or a spike in the clock frequency. Consider a discretized version of the Allan variance~(\ref{oldAllan}),
\begin{align}
\sigma_y^2(\tau) = \frac{1}{2 (p-1)}\sum\limits_{i=1}^{p-1}\left(\bar y_{i+1} - \bar y_i \right)^2\,.
\end{align}
If the phase jump $\Delta \varphi$ occurs at $i=k$, then $\bar y_k = \Delta \varphi / (2\pi \nu_1 \tau)$ and the DM-induced part of the Allan variance is
\begin{align}
\sigma_y^2(\tau) &= \frac{1}{2(n-1)}\left[ \left(\bar y_{k+1} - \bar y_k \right)^2+ \left(\bar y_{k} - \bar y_{k-1} \right)^2\right] \nonumber\\
&= \frac{\Delta \varphi^2}{\tau (T-\tau)(2\pi \nu_1)^2}\,.
\end{align}
If the total Allan deviation is given by $\sigma_y(\tau)=\sigma_0 \tau^{-1/2}$ (typical case of short-term stability), then the DM contribution to $\sigma_0^2$ is 
\begin{align}
(\Delta \sigma_0)^2 = \frac{\Delta \varphi^2}{(2\pi \nu_1)^2 T}\,. \label{phshift}
\end{align}
As one can see, the effect of the phase jump is equivalent to the white noise~\cite{Allan} that can be expected from the Fourier image of the delta-function and was studied before in relation to the atomic clocks~\cite{Greenhall}. Such effect would slightly shift the position of the $\sigma-\tau$ curve, without changing its slope, see Fig.~\ref{pic3}. We suggest to measure the uncertainty in the position of the curve by performing clock comparison in many separate sessions, assuming that the interaction with a DM object happens at least ones in the mission lifetime, but not more frequent than once per several measurement sessions. While the phase jump by itself can be caused by various factors (their nature is still not fully understood even for GPS Rb clocks~\footnote{Dmitry Duev, personal communication.}) and one would need to study a correlation in the responses of several clocks to make sure it was triggered by DM, our method can be used to limit the strength of the DM couplings.
If the average time between DM object interactions is $\TT$, then one needs to conduct $N=\TT / T$ measurement sessions to experience the interaction. Assuming the standard deviation for $\sigma_0$ from $N$ measurement sessions is $\delta \sigma_0 \ll \sigma_0$, we get $\Delta \sigma_0 < (2 N \sigma_0 \delta \sigma_0)^{1/2}$. The value of the phase shift will be given by $\Delta \varphi = 2 \pi \nu_1 \KK_{(n)} \bar \phi^n t_{int}$, $t_{int}\sim 1/(\vv m_\phi)$, where $n$ is defined in Eq.~(\ref{Lagrangian}) and $\KK_{(n)}$ is the response coefficient from Eq.~(\ref{freqshift}), such as $\delta \nu / \nu = \KK_{(n)} \phi^n$. The typical mass of DM object, $M_{mon}$, and an average distance $\bar L$ between them can be estimated from geometric arguments,
\begin{align}
M_{mon} \sim \frac{m_\phi^2 \bar \phi^2}{2}\cdot \frac{1}{m_\phi^3}\sim \rho_{DM} \bar L^3,\quad \bar L^3 = \frac{\TT \vv}{m_\phi^2} \label{geometric}
\end{align}
where we omit numerical factors to make order of magnitude estimates. This gives us the average value of the scalar field inside the DM object,
$\bar \phi^2 = \TT \vv \rho_{DM}/m_\phi$, which can be shown to be the same for topological defects of other dimensionality~\cite{Derevianko:2013oaa}. This value can be further substituted in Eq.~(\ref{phshift}) and leads to
\begin{align}
\KK_{(n)} < \frac{\sigma_0^{1/2} \delta\sigma_0^{1/2} m_\phi^{n/2 + 1}}{\TT^{(n-1)/2}\, \vv^{n/2 - 1} \rho_{DM}^{n/2}}\,.\label{genlimit}
\end{align}
In order to estimate the sensitivity of the method, let us consider a case of linear coupling, $n=1$, and an optical clock, $\KK_{(1)} = K_\alpha/\LL{\gamma}{1}$ (this can be also a microwave clock with all scales but $\LL{\gamma}{1}$ set to infinity). We can further assume a reasonable case $\delta \sigma_0 < 0.1\,\sigma_0$ (or even more conservative $\delta \sigma_0 \sim \sigma_0$) to get the order of magnitude limit
\begin{align}
\LL{\gamma}{1} > \frac{|K_\alpha| \rho_{DM}^{1/2}}{\sigma_0 \vv^{1/2} m_\phi^{3/2}}\,.
\end{align}
As two practical examples,
\begin{align}
\LL{\gamma}{1}^{\mathrm{ACES}} >  10^4 \left(\frac{1\,\mathrm{eV}}{m_\phi} \right)^{3/2}, \quad \LL{\gamma}{1}^{\mathrm{Sr}} > 10^5 \left(\frac{1\,\mathrm{eV}}{m_\phi} \right)^{3/2}\,,\label{aceslimits}
\end{align}
see Fig.~\ref{exclusions2}. In the case of ACES or any other future clock on ISS, $\tau \ll T\sim 3$ min. Putting several clocks in a more distant orbit would increase $T$, lower minimal $m_\phi$ and allow for a long-time common-view comparison with several ground clocks helping to identify the DM origin of the phase jump. Putting a precise atomic clock on a deep-space spacecraft, such as 100 AU mission~\cite{Buscaino:2015fya} and comparing the clock frequency with a less DM-sensitive clock on ground would allow us to study very large DM inhomogeneities and hence very small $m_\phi$.

\begin{figure*}[t]
\centering
\subfigure[\label{Ln}]{\includegraphics[width=7cm]{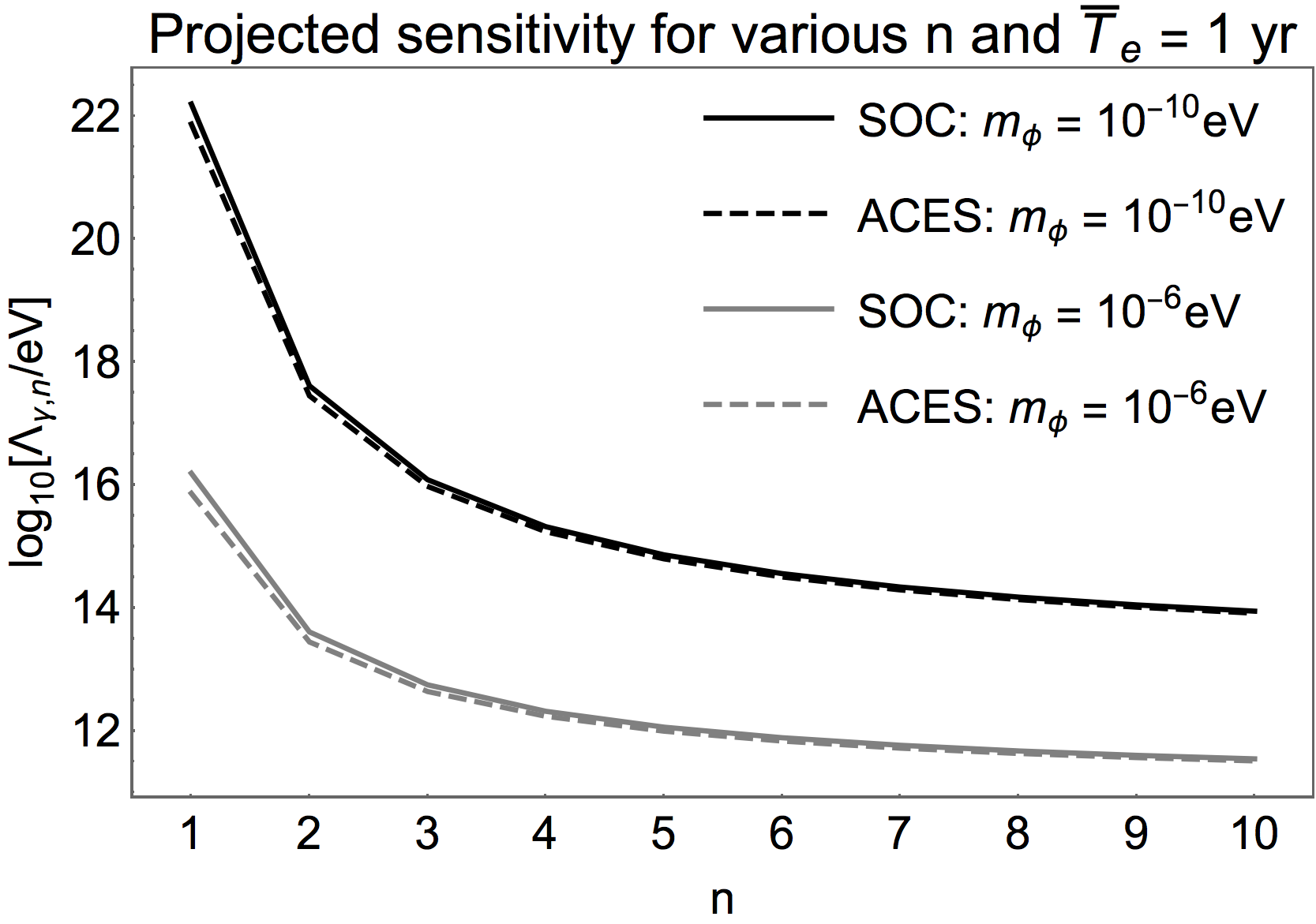}}\hspace{1cm}
\subfigure[\label{apriori}]{\includegraphics[width=7cm]{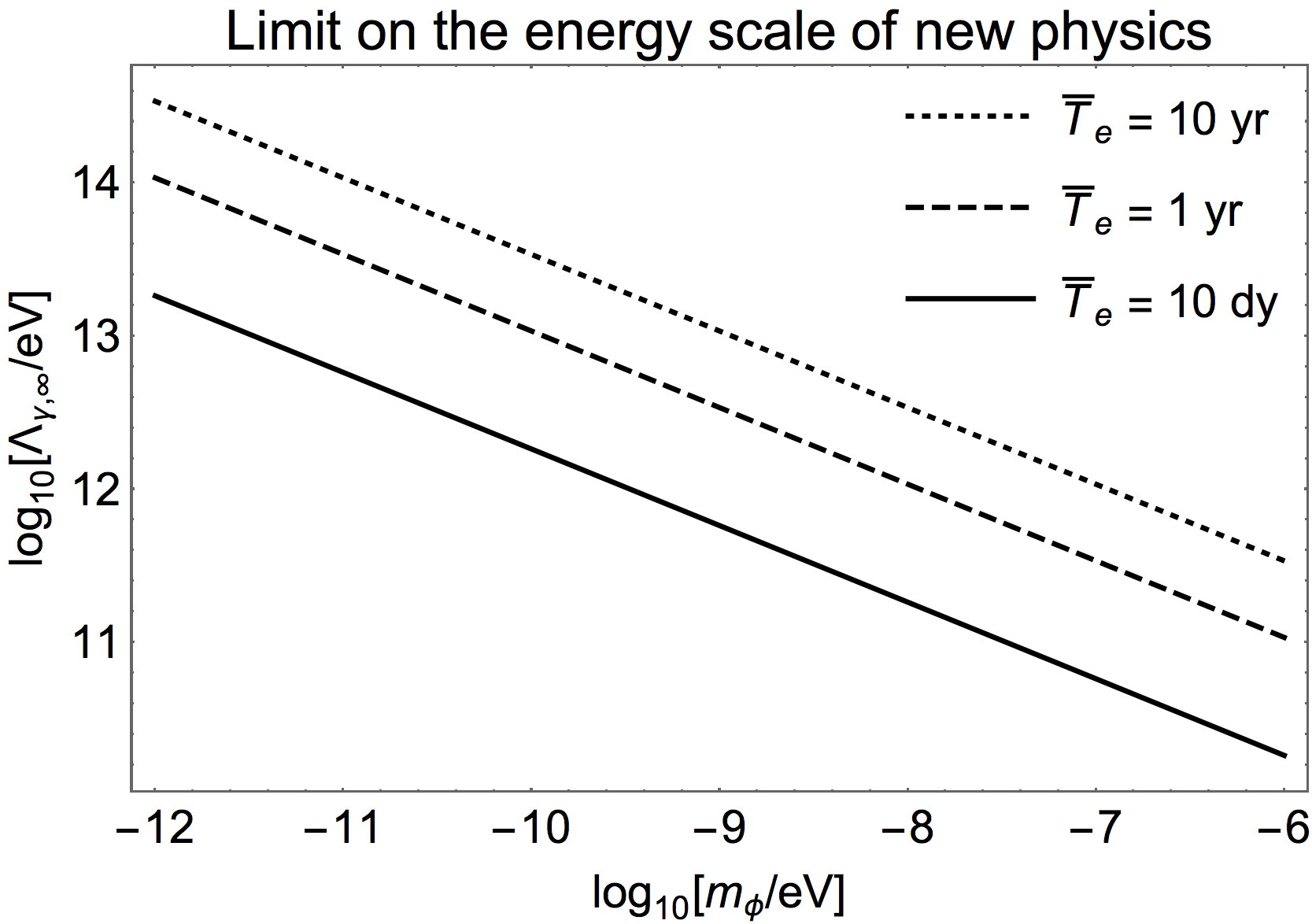}}
\caption{(a) Projected sensitivities for $\LL{\gamma}{n}$ at different $n$ in assumption $\TT=$1 yr and standard deviation for $\sigma_0$ less than 10\%. (b) {\it A priori} lower bounds on the energy scale of new physics allowing for the topological configurations at low energies. \label{anyn}}
\end{figure*}

As we see, for the $n=1$ case $\TT$ drops out from the calculation, and $T$ drops out for any $n$. For $n > 1$ one has to introduce additional assumptions on the value of $\TT$, which makes the limit less robust. As a reference number, one can take $\TT = 1$ yr, which is an approximate duration of the mission. 
However, nothing prevents this number to take any other arbitrary values. For the $n=2$ case we get
\begin{align}
\LL{\gamma}{2} > \frac{|K_\alpha|^{1/2}\TT^{1/4}\rho_{DM}^{1/2}}{\sigma_0^{1/4} \delta\sigma_0^{1/4}m_\phi}\,,\label{oursensitivity}
\end{align}
and, therefore (if $\delta\sigma_0 \sim 0.1 \sigma_0$),
\begin{align}
\LL{\gamma}{2}^{\mathrm{ACES}} >  10^6 \left(\frac{1\,\mathrm{eV}}{m_\phi} \right), \quad \LL{\gamma}{2}^{\mathrm{Sr}} > 2 \times 10^6 \left(\frac{1\,\mathrm{eV}}{m_\phi} \right)\,,\label{aceslimits2}
\end{align}
see Fig.~\ref{exclusions4}. These should be understood as order of magnitude estimates only. One can notice that our sensitivity (\ref{oursensitivity}) has the same dependence on the clock stability, mass $m_\phi$, response coefficient $K_\alpha$ and density $\rho_{DM}$, when compared to Eq.~(\ref{sensitivity}). However, it does not depend on the averaging time $\tau$ and the distance between clocks $l$, and has a weaker dependence on the time between interactions $\TT$. 

Our method does not require that both clocks will interact with the same topological defect, which is an advantage in the case of monopoles and strings. This is also a difference between our approach and the one implemented in Ref.~\cite{Wcislo:2016qng}, where two co-located clocks are sharing the same optical cavity and two readouts are cross-correlated in order to extract the common signal (which is some type of noise by itself) from the instrumental noise. In other words, instead of looking for the {\it similar} contribution to the noise of each clock, we are looking for the {\it difference} in two clocks' noise levels. The conceptual difference between our procedure and the one in Ref.~\cite{Wcislo:2016qng}, naturally, leads to a different functional form of the limit on $\LL{\gamma}{2}$.

\subsection*{Case $1/m_\phi \gg T$}
Assuming, again, that DM exists in a form of a wave and noticing that the integrand in Eq.~(\ref{timeAllan}) can be considered constant in this case, the DM-induced Allan variance is
\begin{align}
\sigma_y^2(t,\tau) = \frac{2 A^2 \sin^4(m_\phi \tau/2)}{(m_\phi \tau/2)^2} \sin^2(m_\phi t + \varphi_0)\,.
\end{align}
Applying further simplifications, we get 
\begin{align}
\sigma_y^2(t,\tau) &\approx \frac{A^2}{2}(m_\phi \tau)^2 \sin(m_\phi t + \varphi_0) \nonumber\\
&= \frac{\KK_{(1)}^2}{2}\rho_{DM}\tau^2 \sin^2(m_\phi t + \varphi_0)\,,
\end{align}
so we see that the effect of DM interaction will manifest itself in additional deviation growing linearly with $\tau$ and oscillating as a function of time, see Fig.~\ref{pic4}. As discussed before, the data for this case should be collected in several separate measurement sessions, and the stability of the clock should be determined for each of them. This is conceptually different from the method of the previous two sections, since we are not searching for the ``bumps'' in the $\sigma-\tau$ diagram but at the variation of the stability curve at large $\tau$ as the function of time, at which the measurement session took place. Let us estimate the sensitivity of an optical clock (or a microwave clock with $\LL{q}{1}, \LL{g}{1} \to \infty$). Taken we know the Allan variance with uncertainty $[\delta \sigma(\tau)]^2$ and no oscillation of $\sigma_y(\tau)$ is observed, we get
\begin{align}
\LL{\gamma}{1} > \frac{|K_\alpha| \rho_{DM}^{1/2} \tau}{2^{1/2}\, \delta \sigma_y(\tau)}\,,
\end{align}
where one can notice no dependence on $m_\phi$. Let us choose $\tau \sim 10\,\mathrm{s} \ll T$ (motivated by the ISS flyby time $T \sim$ few min) and $\delta \sigma_y(\tau) = 0.1\, \sigma_y(\tau)$, then
\begin{align}
\LL{\gamma}{1}^{\mathrm{ACES}} > 4\times 10^{28}\mathrm{eV} , \quad \LL{\gamma}{1}^{\mathrm{Sr}} > 6\times 10^{28}\mathrm{eV} \,.\label{aceslimits1}
\end{align}
Even though these limits are beyond the Planck mass, they are significantly weaker than currently existing limits coming from the Dy spectroscopy and Rb/Cs clock comparison, see Fig.~\ref{exclusions1}. In order to put new limits, one could perform comparison of two clocks with significantly different coefficients $K_\alpha$ in a laboratory with much larger $\tau$. Because this method does not seem to have an advantage comparing to other existing methods, we do not consider limits on $\LL{\gamma}{2}$.

\section{Topological defects with $n>2$}

In case of monopole/string DM, it is useful to consider Eq.~(\ref{genlimit}) with $\KK_{(n)}=K_\alpha/\LL{\gamma}{n}^n$ at any $n$, see Fig.~\ref{Ln}. As one can see, the values $\LL{\gamma}{n}$ are bounded from below by a constant
\begin{align}
\LL{\gamma}{\infty} = \lim\limits_{n \to \infty} \LL{\gamma}{n} = \frac{\TT^{1/2} \vv^{1/2}\rho_{DM}^{1/2}}{m_\phi^{1/2}}\,,\label{LLinf}
\end{align}
so the scale $\Lambda_\gamma$ can not be made indefinitely small with large $n$, see Fig.~\ref{apriori}. The value (\ref{LLinf}) does not depend on $K_\alpha$, $\sigma_0$ or $\delta \sigma_0$ and, therefore, provides us with an {\it a priori} lower limit on the energy scale of new physics that allows for low-dimensional configurations at low energies. The physical meaning of this scale is evident from Eq.~(\ref{geometric}), $\LL{\gamma}{\infty} = \bar \phi$, i.e., if the field strength inside the DM object is comparable to the typical energy scale of the UV-completion of the theory, then the power expansion leading to the Lagrangian Eq.~(\ref{Lagrangian}) breaks down and the effective theory is not valid. 

\section{Conclusions and discussion}
In this paper, we present the analysis of various DM effects on the clock stability measurements. While the stability analysis should not be considered as a general method for DM detection with clocks, in certain cases, it would be preferred, such as when the DM effect resembles a certain type of noise, e.g., the white noise due to interaction with compact DM objects. It is also interesting that the fast periodic variations ($f$ larger than inverse sampling time) will manifest themselves as secondary ``bumps'' in the $\sigma-\tau$ diagram, while the peak in the power spectrum will be missed (due to Nyquist-Shannon-Kotelnikov theorem). Indeed, as can be seen in Figs.~\ref{exclusions1} \& \ref{exclusions3}, there are no existing limits obtained with the spectral analysis for $m_\phi > 10^{-15}$ eV. As a practical matter, most of the existing clock comparison data are presented in the form of Allan variance plots. This makes it possible to extract certain DM limits for the most precise clock comparisons to date.

The main results of our investigation are summarized in Figs.~\ref{exclusions} \& \ref{quadexclusions}. The dashed lines represent sensitivities of the future ISS experiments as well as ``the best case scenario'' thorium nuclear clock. Solid lines correspond to the limits drawn from experimental data, such as the Dy spectroscopy~\cite{VanTilburg:2015oza}, Cs/Rb clock comparison~\cite{Hees:2016gop} and the $\mathrm{Hg}^+/\mathrm{Al}^+$ clock comparison (this paper). The existing cosmological exclusion regions are also presented. The $\mathrm{Hg}^+/\mathrm{Al}^+$ clock comparison allows us to put first limits on the DM coupling in the region $m_\phi > 10^{-15}$ eV, that previously was treated as inaccessible due to the typical clock loop time of order of one or few seconds.
Regarding the monopole/string scenario, if implemented in the future, this will be the first such limit with the linear coupling for $10^{-11}\,\mathrm{eV} < m_\phi < 10^{-6}\,\mathrm{eV}$ for $n=1$ and  $10^{-9}\,\mathrm{eV} < m_\phi < 10^{-6}\,\mathrm{eV}$ for $n=2$ (the upper limit assumes the size of the clock of order of 1 m). Our sensitivity can be underestimated (if the error in $\sigma_0$ determination is below 10\%) as well as overestimated (if the microwave/optical link with ISS introduces additional uncertainties).

Finally, we should comment that we chose the value  $\rho_{DM}$ to represent the energy density of the scalar field to be able to compare the sensitivity of our method to the existing or future methods appearing in the literature. However, the obtained limits are not intended to be understood in the absolute sense, without taking into account all additional assumptions on the properties of the scalar field distribution and its dynamics. Field $\phi$ can be of any nature, not necessarily the DM and, hence, can have a much larger (or much lower) local energy density within the solar system.
This leads us to an additional outlook -- direct detection of new light scalar fields of unknown nature with properties different from DM.

\section*{Acknowledgements}
The authors are grateful to Slava Turyshev, Eric Burt, Jason Williams, Dmitry Duev, Andrei Derevianko and Peter Wolf for useful comments and suggestions. This work was performed at the Jet Propulsion Laboratory, California Institute of Technology, under a contract with the National Aeronautics and Space Administration. \textcopyright~2017 California Institute of Technology. Government sponsorship acknowledged.

\end{document}